\documentclass[floatfix,
reprint,
amsmath,amssymb,
aps,nofootinbib,prd
]{revtex4-2}

\usepackage{subfig}
\usepackage{graphicx}
\usepackage{dcolumn}
\usepackage{bm}
\usepackage{graphicx}	
\usepackage{amsmath}	
\usepackage{mathtools}  
\usepackage{xcolor}
\usepackage{comment}
\usepackage{hyperref}
\usepackage{comment}
\usepackage{tabularx}

\usepackage{caption}
\usepackage{academicons}

\newcommand{\orcidicon}{\includegraphics[width=0.32cm]{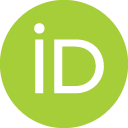}}
\newcommand{\orcid}[1]{\href{https: //orcid.org/#1}{\orcidicon}}

\newcommand{\citelink}[2]{\hyperlink{cite.#1}{#2}}

\usepackage[separate-uncertainty = true,multi-part-units=single]{siunitx}

\usepackage{listings}
\lstdefinestyle{mystyle}{
    backgroundcolor=\color{backcolour},   
    commentstyle=\color{codegreen},
    keywordstyle=\color{magenta},
    numberstyle=\tiny\color{codegray},
    stringstyle=\color{codepurple},
    basicstyle=\ttfamily\footnotesize,
    breakatwhitespace=false,         
    breaklines=true,                 
    captionpos=b,                    
    keepspaces=true,                 
    showspaces=false,                
    showstringspaces=false,
    showtabs=false,                  
    tabsize=2
}
\usepackage{realboxes}
\usepackage{enumitem}
\usepackage{placeins}

\usepackage{tikz}
\usepackage{tikz-3dplot}
\usetikzlibrary{angles, quotes} % Load the angles and quotes libraries
\usepackage{Functions_Sphere}
\usetikzlibrary{arrows, decorations.markings}
\usetikzlibrary{arrows.meta}
\tikzset{viewport/.style 2 args={
    x={({cos(-#1)*\RadiusSphere cm},{sin(-#1)*sin(#2)*\RadiusSphere cm})},
    y={({-sin(-#1)*\RadiusSphere cm},{cos(-#1)*sin(#2)*\RadiusSphere cm})},
    z={(0,{cos(#2)*\RadiusSphere cm})}
}}

\usepackage{listings}
\lstdefinestyle{mystyle}{
    backgroundcolor=\color{backcolour},   
    commentstyle=\color{codegreen},
    keywordstyle=\color{magenta},
    numberstyle=\tiny\color{codegray},
    stringstyle=\color{codepurple},
    basicstyle=\ttfamily\footnotesize,
    breakatwhitespace=false,         
    breaklines=true,                 
    captionpos=b,                    
    keepspaces=true,                 
    showspaces=false,                
    showstringspaces=false,
    showtabs=false,                  
    tabsize=2
}
\usepackage{realboxes}
\usepackage{enumitem}
\usepackage{placeins}

\usepackage{tikz}
\usepackage{tikz-3dplot}
\usetikzlibrary{angles, quotes} % Load the angles and quotes libraries
\usepackage{Functions_Sphere}
\usetikzlibrary{arrows, decorations.markings}
\usetikzlibrary{arrows.meta}
\tikzset{viewport/.style 2 args={
    x={({cos(-#1)*\RadiusSphere cm},{sin(-#1)*sin(#2)*\RadiusSphere cm})},
    y={({-sin(-#1)*\RadiusSphere cm},{cos(-#1)*sin(#2)*\RadiusSphere cm})},
    z={(0,{cos(#2)*\RadiusSphere cm})}
}}

\def\RadiusSphere{2}

\definecolor{darkgreen}{rgb}{0.0, 0.5, 0.0}

\begin{document}
	\title[Enhanced Disruption of Axion Minihalos by Multiple Stellar Encounters in the Milky Way]{Enhanced Disruption of Axion Minihalos by Multiple Stellar Encounters in the Milky Way}
	
	\author{Ian DSouza\orcid{0009-0004-3142-3898}}
	\email{ids29@uclive.ac.nz}
	\author{Chris Gordon\orcid{0000-0003-4864-5150}}
	\email{chris.gordon@canterbury.ac.nz}
	\author{John C. Forbes\orcid{0000-0002-1975-4449}}
	\email{john.forbes@canterbury.ac.nz}
	\affiliation{
		School of Physical and Chemical Sciences, University of Canterbury,
		New Zealand
	}

	\date{\today}
	
	\begin{abstract}
If QCD axion dark matter formed post-inflation, axion miniclusters emerged from isocurvature fluctuations and later merged hierarchically into minihalos. These minihalos, potentially disrupted by stellar encounters in the Milky Way, affect axion detectability. We extend prior analyses by more accurately incorporating multiple stellar encounters and dynamical relaxation timescales, simulating minihalo orbits in the Galactic potential.

Our results show stellar interactions are more destructive than previously estimated, reducing minihalo mass retention at the solar system to ~30\%, compared to earlier estimates of ~60\%. This enhanced loss arises from cumulative energy injections when relaxation periods between stellar encounters are accounted for.

The altered minihalo mass function implies a larger fraction of axion dark matter occupies inter-minihalo space, potentially increasing the local axion density and improving haloscope detection prospects. This work highlights the significance of detailed modeling of stellar disruptions in shaping the axion dark matter distribution.

\end{abstract}
	
	\maketitle

\section{Introduction}

The axion, originally proposed as a solution to the strong CP problem in quantum chromodynamics \cite{Peccei:1977hh, peccei1977constraints, Weinberg1978, Wilczek1978, Kim1979, Zhitnitsky1980, ShifmanVainshteinakharov1980, DineFischlerSrednicki1981, KimCarosi2010}, has emerged as a compelling dark matter candidate
(see \cite{IrastorzaRedondo18, DiLuzio20, Chadha-Day21, Adams22, O'Hare24} for recent reviews). 
If the symmetry breaking leading to the birth of axions occurs after inflation, then the dynamics of axion strings and domain walls play a crucial role in shaping the axion relic abundance and distribution, as demonstrated in numerical simulations \cite{Fleury2016,Klaer2017, Gorghetto2018, Buschmann2020, Gorghetto2021, Buschmann2022, OHare2022}. These processes set the stage for the formation of axion miniclusters, which arise from isocurvature perturbations on scales set by the QCD phase transition \cite{Hogan1988, Kolb1993, KolbTkachev1994, KolbTkachev1996, ZurekHoganQuinn07}. 
 
 The hierarchical merging of miniclusters leads to axion minihalos, typically with masses of $10^{-8} M_{\odot}$ today. Previous studies have explored the formation and evolution of axion minihalos, including their mass function and spatial distribution \cite{Hogan1988, Kolb1993, KolbTkachev1994, KolbTkachev1996, ZurekHoganQuinn07, Hardy16, DavidsonSchwetz16, Enander2017, FairbairnMarshQuevillon2017, Fairbairn2018, Eggemeier2019, BlinovDolanDraper2020, Eggemeier20, Croon20, X2021, Edwards21, Ellis2022, Dandoy2024}. The presence of minihalos can impact observational signatures such as axion direct detection experiments using haloscopes, particularly
as the earth may be in an axion minihalo void, which would make detection more difficult\cite{Eggemeier23,ohareAxionMiniclusterStreams2024}.

One key aspect affecting the survival and distribution of axion minihalos is their interaction with stars in the Milky Way galaxy. Stellar encounters can inject energy into minihalos, leading to mass loss \cite{Berezinsky2013, Tinyakov2016, DokuchaevEroshenkoTkachev17, Kavanagh21, S2024, ohareAxionMiniclusterStreams2024, dsouza2024}. Accurately modeling these encounters is essential for predicting the present-day abundance and properties of minihalos.

Recent work by Ref.~\cite{S2024}, hereafter referred to as \citelink{S2024}{S2024}, examined the disruption of axion minihalos due to stellar encounters, employing a linear addition of energy injections from multiple encounters. However, as pointed out in Ref.~\cite{dsouza2024} (hereafter referred to as \citelink{dsouza2024}{Paper 1}), this approach may underestimate the cumulative effects when minihalos have time to relax between encounters\footnote{There is some overlap between the authors of the current paper and \citelink{dsouza2024}{Paper 1}.}.

In this paper, we account for the dynamical timescales of minihalos to determine whether they can relax between encounters, leading to a nonlinear addition of energy injections when appropriate. By generating a population of minihalo orbits using Monte Carlo simulations and evolving them within a model of the Galactic potential, we compute the stellar-disrupted mass function of minihalos more precisely.

The paper is organized as follows. In Sec.~\ref{sec: pre-infall mass function}, we discuss the pre-infall mass function of axion minihalos formed from isocurvature perturbations. Sec.~\ref{sec: Formation halos from adiabatic perturbations} describes the formation of larger adiabatic halos from adiabatic perturbations. In Sec.~\ref{sec: Formation halos from adiabatic perturbations}, we derive the undisrupted mass function of minihalos within adiabatic halos. The mass-concentration relationship for minihalos is presented in Sec.~\ref{sec: Mass-Concentration relationship}. In Sec.~\ref{sec: Accounting for multiple stellar encounters}, we detail our method for accounting for multiple stellar encounters, considering the minihalo's ability to relax between encounters.

We then describe our Monte Carlo simulations in Secs.~\ref{sec: Monte Carlo sampling of orbits in singular isothermal sphere} to \ref{sec: Monte Carlo Simulations to determine the stellar-disrupted mass function of minihalos}. Finally, we present our results on the stellar-disrupted mass function and discuss the implications for axion dark matter detection in Sec.~\ref{sec: computing the survival fraction of the minihalo} and give our conclusions in Sec.~\ref{sec: conclusions}.

	\section{Pre-infall mass function}\label{sec: pre-infall mass function}
	 The axion isocurvature perturbations gravitationally collapse to form axion miniclusters around matter-radiation equality. These initially formed axion miniclusters undergo hierarchical mergers to form larger minihalos. The comoving number density of minihalos in a given mass range is quantified by the \textit{mass function} of minihalos.
	Ref.~\cite{X2021}, hereafter referred to as \citelink{X2021}{X2021}, performed numerical simulations to generate the mass function of such minihalos. They later fitted a modified  Sheth-Tormen formula \cite{ShethTormen99} to match the results of their simulations. We use this formula for the mass function of axion minihalos.
	
	The isocurvature growth function tells us how the isocurvature density fluctuations of axions evolve with redshift. We use the Code for Anisotropies in the Microwave Background (\texttt{CAMB}; \cite{lewis2000efficient, howlett2012cmb}) package in Python to determine this growth function. We assume a flat $\Lambda$CDM cosmology with $\Omega_{\rm m} = 0.2814$, $\Omega_\Lambda = 0.7186$, scalar spectral index $n_{\rm s} = 0.9667$, and $h = 0.697$. These are the values that \citelink{S2024}{S2024} assumed and they are consistent with the Planck 2018 Results \cite{aghanim2020planck}. We also set $\Omega_{\rm r} = 8.6113\times10^{-5}$, $\Omega_{\rm b}h^2 = 0.0240$, $\Omega_{\rm m}h^2=0.1404$, and $\sigma_{\rm 8} = 0.796$ from the Planck 2018 results.
	We set $T_{\rm CMB} = 2.7255\mathrm{K}$ from Ref.~\cite{Fixsen2009}.
	Using \texttt{CAMB}, we can set up the initial conditions of isocurvature perturbations for cold dark matter. We initially calculate the value of the growth function relative to redshift $z=100$, which is deep in the matter-dominated era.
	We do this by first calculating the power spectrum at the desired redshift and also at $z=100$. We then calculate the square of the growth function as the ratio of the value of the power spectrum at the desired redshift to the value of the power spectrum at $z=100$. When selecting the value of the power spectrum, we look at the power spectrum corresponding to small length scales (or high $k$) because it is in this regime that the power spectrum becomes scale-independent. 
	In our code, we consider the $k$ values up to $2h\,\mathrm{Mpc}^{-1}$.

	We use \texttt{CAMB} to evolve the isocurvature perturbations. See Appendix \ref{app: renormalizing the CAMB growth function} for details.
	Knowing how the isocurvature perturbations of axion dark matter evolve, we can now calculate the mass function $\frac{\mathrm{d} n_0}{\mathrm{d} M}\left(M, z\right)$ at a given redshift ($z$) and mass ($M$) using a modified Sheth-Tormen formalism. See Appendix~\ref{app: pre-infall mass function} for further details.

\section{Formation of halos from adiabatic perturbations}
\label{sec: Formation halos from adiabatic perturbations}
Sec.~\ref{sec: pre-infall mass function} detailed how axion minihalos are formed from isocurvature perturbations. On the other hand, there exist adiabatic perturbations in the axion density field in the primordial universe as well.
These perturbations collapse to form larger halos, which we refer to as {\em adiabatic halos\/}. Note that  \citelink{X2021}{X2021} and \citelink{S2024}{S2024} refer to them as {\em CDM halos\/}.
These halos generally form much later than the axion minihalos and can be host to galaxies and galaxy clusters.

We use the \texttt{hmf} package \cite{hmf}
to compute the collapse fraction of adiabatic halos, $f_{\rm adiab}$, as a function of redshift. 
We generate the redshift-dependent mass function $\mathrm{d}n^{\rm adiab}/\mathrm{d}M$  using the Press-Schecter formula \cite{PressSchecter74} and set the growth model to ``CambGrowth" and the transfer model to ``CAMB" in the \texttt{hmf} package. This mass function incorporates both the baryonic and cold dark matter. When the minihalos are in the adiabatic halos, they are predicted to freeze in their evolution due to the high virial velocities. This will only happen for adiabatic halos that are substantially more massive than the minihalos.
To account for this, \citelink{S2024}{S2024} states that they impose a minimum adiabatic mass halo of  $M_{\rm min} = 10^{-2} M_\odot$. They base their computation on the results of \citelink{X2021}{X2021}, which states that they impose $M_{\rm min} = 10^{-3} M_\odot$. However, 
they 
actually used  $M_{\rm min} = 10^2 M_\odot$ 
\cite{Xiao2023Communication}.    
To ascertain how sensitive the result is to this choice, we used two different values of $M_{\rm min}$ in all our calculations.
We choose an upper bound for the adiabatic halos of  $M_{\rm max}=10^{20} M_\odot$ because the mass function of adiabatic halos as generated by the \texttt{hmf} package is sufficiently suppressed for masses $> 10^{20} M_\odot$, and increasing the value of $M_{\rm max}$ doesn't affect the value of the collapse fraction. We  calculate the collapse fraction 
using
\begin{equation}\label{eq: collapse fraction}
	f_{\rm adiab}(z) = \frac{1}{\bar{\rho}_{\rm m}} \int_{M_{\rm min}}^{M_{\rm max}} M \frac{\mathrm{d}n^{\rm adiab}}{\mathrm{d}M}(M, z) \mathrm{d}M \ ,
\end{equation}
where $M$ is the mass of the adiabatic halo and $\bar{\rho}_{\rm m}$ is the average comoving mass-density of matter (which is also equal to the average physical mass-density of matter today). We numerically perform the integral in Eq.~(\ref{eq: collapse fraction}) using Simpson's rule from the \texttt{SciPy} package.

\begin{figure}
	\includegraphics[width=\columnwidth]{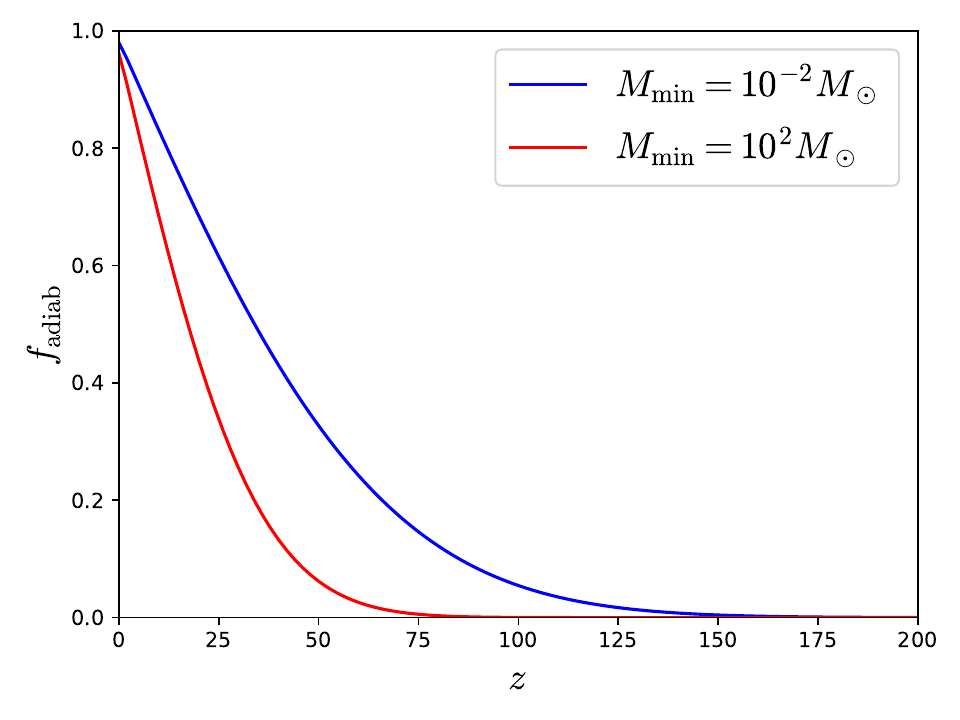}
	\caption{The collapse fraction of adiabatic halos, with masses between $M_{\rm min}$ and $M_{\rm max}=10^{20}M_\odot$, plotted against redshift. %
	}
	\label{fig: collapse fraction}
\end{figure}

We compute the collapse fraction using Eq.~(\ref{eq: collapse fraction}) for different values of redshift. Fig.~\ref{fig: collapse fraction} shows how the collapse fraction of adiabatic halos changes with redshift. 

We further numerically compute the derivative $\mathrm{d}f_{\rm adiab} / \mathrm{d}z$.
We then create an interpolation object that takes as input the redshift and outputs the value of $\mathrm{d}f_{\rm adiab} / \mathrm{d}z$ corresponding to that redshift.

%\section{Undisrupted mass function}\label{sec: undisrupted mass function}
\citelink{S2024}{S2024} gives the expression of the undisrupted mass function of minihalos $\mathrm{d} n_{\mathrm{f}} / \mathrm{d} M$ inside the adiabatic halos at redshift $z=0$ as: 
\begin{equation}\label{eq: undisrupted mass function}
	\frac{\mathrm{d} n_{\mathrm{f}}}{\mathrm{d} M}(M)=\int_{z_{\mathrm{eq}}}^0 \mathrm{d} z \frac{\mathrm{d} f_{\rm adiab}}{\mathrm{d} z}\left(z\right) \frac{\mathrm{d} n_0}{\mathrm{d} M}\left(M, z\right),
\end{equation}
where $\mathrm{d} n_0 / \mathrm{d} M$ is the pre-infall mass function of minihalos, $z$ is the infall redshift of the minihalos, $z_{\rm eq}$ is the redshift of matter-radiation equality.

\section{Mass-Concentration relationship}\label{sec: Mass-Concentration relationship}
We consider minihalos that have the spherically symmetric Navarro–Frenk–White (NFW) density profile \cite{NFW1997universal}: 
\begin{equation}
	\rho_{\rm NFW}(r) = \frac{\rho_{\rm s}}{\frac{r}{r_{\rm s}} \left(1 + \frac{r}{r_{\rm s}}\right)^2}\ ,
\end{equation}
where $r$ is the distance from the center of the minihalo, $r_{\rm s}$ is called the scale radius, and $\rho_{\rm s}$ is called the scale density. Note that when the axion miniclusters initially collapse from isocurvature density fluctuations, they do not have an NFW profile. However, by the time the adiabatic halos form, these primordial miniclusters will undergo hierarchical merging to give rise to NFW density minihalos \citelink{X2021}{X2021}. The virial radius ($r_{\rm vir}$) of the NFW density profile minihalo at a given redshift ($z$) is defined
as that radius inside which the mean density of the minihalo is given by
\begin{equation}\label{eq:virial density}
	\bar{\rho}_{\mathrm{vir}}(z) = 200 \,\rho_{\rm crit}(z)\ ,
\end{equation}
and  $\rho_{\rm crit}$ is the cosmological critical density at redshift $z$. 
The virial mass $M $ of the minihalo is the mass enclosed within the virial radius. In this article, we sometimes drop the term \textit{virial} and simply refer to it as the mass of the minihalo. 
The concentration parameter of the minihalo is defined as $c\equiv r_{\rm vir}/r_s$.
For a given minihalo, the concentration, virial mass, and the redshift at which they are evaluated are related to each other \cite{lee2021probing, bullock2001profiles}. 
\citelink{S2024}{S2024} has made available \cite{ShenCode} (and presented in Fig.~1 in their article) a tabulated relationship between these quantities for three different axion masses ($m_{\rm a} = 1.25, 25, 500 \mu\mathrm{eV}$). We adopt this relationship in our calculations.

\section{Accounting for multiple stellar encounters}
\label{sec: Accounting for multiple stellar encounters}
In \citelink{dsouza2024}{Paper 1}, we addressed the mass loss incurred by a minihalo during an interaction with a star. In this article, we would like to extend that analysis to the mass loss incurred by a population of minihalos in the  
Milky Way galaxy.

Each time a minihalo has a stellar encounter, there is an associated energy injection, which we quantify as 
\begin{equation}\label{eq: E_frac definition}
	E_{\rm frac} \equiv \frac{\Delta E}{E_{\rm bind}}\ ,
\end{equation}
where $\Delta E$ is the total injected energy into the minihalo within its virial radius.  $E_{\rm bind}$ is the binding energy of the minihalo inside the virial radius.

Let's say there are $N$ stellar encounters by a minihalo. Thus, we have $N$ single encounter events, which we would now like to approximate as an effective single encounter event with an effective energy injection parameter $E_{\rm frac, eff}$, such that the survival fraction of the minihalo, in either case, is the same. 

Ref.~\cite{stucker2023effect} proposed the following formula to evaluate the effective energy injection parameter corresponding to multiple stellar encounters: 
\begin{equation}\label{eq: Stucker et al's expression for effective energy injection parameter}
	E_{\text {frac}, \text {eff}}=\left(\sum_{i=1}^N E_{\text {frac}, i}^{p / 2}\right)^{2 / p} \ .
\end{equation}

When two stellar encounters happen in quick succession such that the minihalo doesn't have enough time to gravitationally relax in between encounters, the minihalo is unable to tell if it has been subjected to two stellar encounters of known energy injection parameters or a single stellar encounter with a higher energy injection parameter. In such a case, what does get added up linearly is the total energy $\Delta E$ injected into the minihalo within its virial radius. Thus, the effective energy injection parameter corresponding to these two single stellar encounters is
\begin{align}\label{eq: effective E_frac for two encounters when relaxation is not allowed in between encounters}
	E_{\rm frac,eff} &= \frac{\Delta E_1 + \Delta E_2}{E_{\rm bind}} \nonumber\\
	&= E_{\rm frac,1} + E_{\rm frac,2} \ ,
\end{align}
where $E_{\rm frac,i} \equiv \Delta E_i / E_{\rm bind}$, for $i=1,2$. Note that in the first equality of Eq.~(\ref{eq: effective E_frac for two encounters when relaxation is not allowed in between encounters}), we did not have a binding energy term separately for each encounter. This is because since the minihalo does not have time to signifcantly change in between encounters, its density profile just before the start of either encounter is approximately the same. If the density profile is the same, so should the binding energy of the minihalo. It is precisely this fact that leads us to the conclusion that the energy injection parameters are added linearly when the minihalo does not have time to change in between encounters. Comparing Eq.~(\ref{eq: effective E_frac for two encounters when relaxation is not allowed in between encounters}) to Eq.~(\ref{eq: Stucker et al's expression for effective energy injection parameter}), we see that $p=2$ when the minihalo doesn't have enough time to change in between encounters.

On the other hand, consider the scenario where we have two stellar encounters with a large amount of time in between encounters. The minihalo is able to completely gravitationally relax in between the encounters. Now, we are no longer able to use the method in Eq.~(\ref{eq: effective E_frac for two encounters when relaxation is not allowed in between encounters}) because the binding energy of the minihalo just before each encounter is different. The amount by which it is different will depend on the energy injected into the minihalo during the first encounter as this will change the density profile of the relaxed minihalo and hence the binding energy of the minihalo just before the second encounter. To address this, in \citelink{dsouza2024}{Paper 1}, we parameter-fit the value of $p$ in  Eq.~(\ref{eq: Stucker et al's expression for effective energy injection parameter}) to various such multiple encounter cases. We found that $p \lesssim 1$ in this case. The smaller the value of $p$ in Eq.~(\ref{eq: Stucker et al's expression for effective energy injection parameter}), the larger is the value of $E_{\text {frac}, \text {eff}}$ for fixed $E_{\text {frac}, i}$. To be on the conservative side, we choose $p=1$. Thus, using Eq.~(\ref{eq: Stucker et al's expression for effective energy injection parameter}) for our two encounter case with complete minihalo relaxation in between encounters, the effective energy injection parameter is given by
\begin{equation}
	E_{\rm frac,eff} =  \left( E_{\rm frac,1}^{1/2} + E_{\rm frac,2}^{1/2} \right)^2
\end{equation}

To add up the energy injection parameters corresponding to two consecutive stellar encounters, we must choose whether we want to add them with $p=1$ or $p=2$. To do this, we first define the dynamical time of the minihalo. This is a quantitative measure of how fast the minihalo is able to relax. The dynamical time chosen could also be called the crossing time. It is defined as the time taken by a small test particle that is released from the surface of the minihalo under the influence of the gravitational potential of the minihalo, to reach the center of the minihalo. This effectively tells us how fast perturbations on the surface of the minihalo propagate through its volume. For simplicity, we consider a homogeneous sphere (instead of the NFW profile of the minihalo) to calculate this crossing time because the crossing time is independent of the radius from which the particle is released in this case. Thus, the dynamical time of the minihalo is given by \cite{binneyTremaine}
\begin{equation}\label{eq: dynamical time}
	t_{\mathrm{dyn}}=\sqrt{\frac{3 \pi}{16 G \bar{\rho}_{\mathrm{vir}}}}
\end{equation}

We compare the time between consecutive stellar encounters to the dynamical time of the minihalo. If the time between consecutive encounters is smaller than the dynamical time, then the minihalo doesn't have enough time to relax and we add up the energy injection parameters  linearly with $p=2$. On the other hand, if the time between consecutive stellar encounters is more than the dynamical time, then the minihalo has enough time to relax and we add up the energy injection parameters non-linearly with $p=1$. We refer to this approach as the {\em hybrid method\/}.

\section{Generating a population of minihalo orbits in the Milky Way galaxy}\label{sec: Monte Carlo sampling of orbits in singular isothermal sphere}

The energy injected into a minihalo during its lifetime in the galaxy can vary depending on the actual orbit of the minihalo since the stellar density varies with position in the galaxy. Thus, we look into generating a large number of orbits and calculate the energy injected into minihalos for each of those orbits. In modeling the population of minihalos, we assume that the minihalos are distributed in the galaxy according to the density profile of a singular isothermal sphere given by the spherically symmetric density profile and potential (\citelink{S2024}{S2024}):   
\begin{equation}
\label{eq:singular isothermal sphere}
	\rho(r)=\frac{V_{\mathrm{C}}^2}{4 \pi G r^2}, \quad \Phi(r)=V_{\mathrm{C}}^2 \ln \left(\frac{r }{ r_0}\right),
\end{equation}
where $V_{\mathrm{C}} = 200\, \mathrm{km / s}$ is the constant circular velocity of the singular isothermal sphere - any minihalo in a circular orbit around the galaxy's center will have this speed, no matter what radius it orbits at:  this emulates the flat rotation curves in our galaxy at sufficiently large radii. Also, $r_0 = 10\, \mathrm{kpc}$ is the reference radius of zero potential. We expect the minihalos to be distributed around the galaxy's center according to an NFW profile. However, the phase space distribution of an NFW profile does not have an analytical form, making it difficult to randomly draw orbits with a given total energy and angular momentum. Thus, as done, for example, by \citelink{S2024}{S2024}, we use the singular isothermal sphere.

We use the Monte Carlo sampling procedure outlined in the appendix of Ref.~\cite{van1999substructure} to sample orbits from a singular isothermal sphere density distribution. We sample such that the minihalo is present in the solar neighborhood today.
This is done by setting the galactocentric radius of observation to be $r_{\rm obs}=8\, \mathrm{kpc}$. 
We assume the phase space distribution function is isotropic. Hence, the term $h(\eta)$ in that article is set to 1. There is a typo in the Monte Carlo procedure outlined in the appendix of Ref.~\cite{van1999substructure}. The condition $\mathcal{R}_4 > P_{\text{comp }}\left(\eta_{\text{try }}\right) / P\left(\eta_{\text{try }}\right)$ should be $\mathcal{R}_4  >   P\left(\eta_{\text{try }}\right) /  P_{\text{comp }}\left(\eta_{\text{try }}\right) $
\cite{vandenBoschCommunication}.
We evaluate the total energy per unit mass $E$ of a randomly drawn orbit as: 
\begin{equation}
	E = \Phi(r_{\rm obs}) - V_{\mathrm{C}}^2 \ln(1 - \mathcal{R}_1)\ ,
\end{equation}
where $\mathcal{R}_1$ is a random number in the interval [0, 1] - see Ref.~\cite{van1999substructure} for details. 
The angular momentum per unit mass $L$ of the randomly drawn orbit is given by: 
\begin{equation}
	L = \eta r_{\rm C}(E) V_{\mathrm{C}}\ ,
\end{equation}
where $\eta$ is called the orbital circularity (related to the orbital eccentricity) and $r_{\rm C}(E)$ is the radius of a circular orbit that has given energy per unit mass $E$. It is given by (\citelink{S2024}{S2024}): 
\begin{equation}
	r_{\rm C}(E)=r_0 \exp \left(\frac{E}{V_{\mathrm{C}}^2}-\frac{1}{2}\right)\, .
\end{equation}

From here on out, we will refer to $E$ and $L$ as just the energy and angular momentum, respectively, of the orbit, but note that they are actually the energy per unit mass and angular momentum per unit mass, respectively.

\section{Initial conditions to evolve an orbit}\label{sec: initial conditions to evolve an orbit}
In the previous section, we sampled orbits from a singular isothermal sphere density distribution such that the minihalos in those orbits are present in the solar neighborhood today. We found the energies and angular momenta of those orbits. We now have to evolve those orbits backward in time to find out the past positions of those minihalos since the stellar density distribution depends on the position, and we want to estimate the mass disruption effects of the stellar population on each minihalo. To do this, we need to convert the energy $E$ and angular momentum magnitude $L$ to the initial state parameters of the minihalo. The state parameters here are the position and velocity vectors of the minihalo with respect to the galactic frame. 
We use the definition of angular momentum of the orbit as: 
\begin{equation}\label{eq: definition of L}
	{ L} = { r}_{\rm obs} \, { v}_{\mathrm{init,}\perp}\ ,
\end{equation}
where $v_{\mathrm{init,}\perp}$ is the magnitude of the initial velocity vector of the minihalo projected onto the plane perpendicular to the radial direction (the line connecting the Galactic center and the minihalo) at the initial position of the minihalo. Using Eq.~(\ref{eq: definition of L}), we can compute the value of $v_{\mathrm{init,}\perp}$. Next, we compute the magnitude of the initial velocity vector using the definition of the total energy $E$ as: 
\begin{equation}\label{eq: definition of E}
	E = \Phi(r_{\rm obs}) + \frac{1}{2}v_{\rm init}^2\ ,
\end{equation}
where $v_{\rm init}$ is the magnitude of the initial velocity vector of the minihalo. Eq.~(\ref{eq: definition of E}) allows us to compute the value of $v_{\rm init}$. Next, we use the vector addition relation: 
\begin{equation}\label{eq: velocity vector addition}
	v_{\rm init}^2 = v_{\mathrm{init,}\parallel}^2 + v_{\mathrm{init,}\perp}^2\ ,
\end{equation}
where $v_{\mathrm{init,}\parallel}$ is the magnitude of the radial component of the initial velocity vector of the minihalo. Using Eq.~(\ref{eq: velocity vector addition}), we can calculate the value of $v_{\mathrm{init,}\parallel}$.

\begin{figure}
    \centering
\begin{tikzpicture}
        \begin{scope}[viewport={160}{15}, very thin, rotate around z=-45,scale=2]

            \draw[dashed] (\ToXYZr{0.5}{90}{45}) -- (\ToXYZr{0.78}{90}{45}); % start of X_axis
            \draw[dashed,color=gray, opacity=0.9] (\ToXYZr{0.78}{90}{45}) -- (\ToXYZr{1}{90}{45}); % middle greyed out part of X_axis
            \draw[dashed,->] (\ToXYZr{1}{90}{45}) -- (\ToXYZr{1.7}{90}{45}) node[anchor=north]{$X$}; % end of X_axis
            
            \draw[dashed,->] (\ToXYZr{0.5}{90}{45}) -- (-0.56568542, 1.27279221, 0) node[anchor=north]{$Y$};
            \draw[dashed,->] (\ToXYZr{0.5}{90}{45}) -- (0.35355339, 0.35355339, 0.75 ) node[anchor=south]{$Z$};
            \node at (0.36769553, 0.36769553, -0.05) {$\mathcal{O}$};

            % Define vertices of the square in the plane perpendicular to the X-axis
            % \coordinate (A) at (\ToXYZr{1.064177772475912}{90+20}{45+20});  % Top-right
            % \coordinate (B) at (\ToXYZr{1.064177772475912}{90-20}{45+20});  % Bottom-right
            % \coordinate (C) at (\ToXYZr{1.064177772475912}{90-20}{45-20});  % Bottom-left
            % \coordinate (D) at (\ToXYZr{1.064177772475912}{90+20}{45-20});  % Top-left
            \coordinate (A) at (0.28284271, 1.13137085, 0.375);  % Top-right
            \coordinate (B) at (0.28284271,  1.13137085, -0.375);  % Bottom-right
            \coordinate (C) at (1.13137085,  0.28284271, -0.375);  % Bottom-left
            \coordinate (D) at (1.13137085, 0.28284271, 0.375);  % Top-left
            
            % Draw the square
            \draw[thick] (A) -- (B) -- (C) -- (D) -- cycle;

            \node at (\ToXYZr{1}{90}{45}) {\Large $\star$};

            % Define the starting point of the vector at $\star$
            \coordinate (Sun) at (\ToXYZr{1}{90}{45});
            
            % Define the end point of the initial velocity vector
            % \coordinate (velocity_vector) at (\ToXYZr{1.3989663259659064}{79.70575598962567}{55.46454366173468}); 
            % % Define the end point of the parallel component of the initial velocity vector
            % \coordinate (velocity_vector_parallel) at (\ToXYZr{1.3535533905932737}{90}{45});
            % % Define the end point of the perpendicular component of the initial velocity vector
            % \coordinate (velocity_vector_perpendicular) at (\ToXYZr{1.0606601717798212}{76.3669777746336}{59.03624346792648});

            % Define the end point of the initial velocity vector
            \coordinate (velocity_vector) at (0.80961941, 1.30459415, 0.35); 
            % Define the end point of the parallel component of the initial velocity vector
            \coordinate (velocity_vector_parallel) at (1.05710678, 1.05710678, 0.);
            % Define the end point of the perpendicular component of the initial velocity vector
            \coordinate (velocity_vector_perpendicular) at (0.45961941, 0.95459415, 0.35);
            
            % Define the end point of the vertical dashed line on the plane
            \coordinate (vertical_dashed_line) at (0.70710678, 0.70710678, 0.35);

            % Draw the vectors
            \draw[thick, ->, blue] (Sun) -- (velocity_vector_parallel) node[anchor=north, xshift=3pt, yshift=0pt] {$\vec{v}_{\mathrm{init,}\parallel}$};
            \draw[thick, ->, darkgreen] (Sun) -- (velocity_vector_perpendicular) node[anchor=north, xshift=6pt, yshift=-10pt] {$\vec{v}_{\mathrm{init,}\perp}$};
            \draw[thick, ->, red] (Sun) -- (velocity_vector) node[anchor=south, xshift=-6pt, yshift=-2pt] {$\vec{v}_{\rm init}$};
            
            % Draw dotted black lines
            \draw[dash pattern=on 0.6pt off 1pt, line width=0.6pt, black] (velocity_vector) -- (velocity_vector_parallel);
            \draw[dash pattern=on 0.6pt off 1pt, line width=0.6pt, black] (velocity_vector) -- (velocity_vector_perpendicular);

            % Draw the vertical dashed line on the plane
            \draw[dashed, black] (Sun) -- (vertical_dashed_line);
            
            % Define the top point above the star
            \coordinate (top_of_plane) at (\ToXYZr{1.3}{90}{45}); % Adjust the radius to position at the top of the plane

            % Draw the dotted line
            \draw[dotted] (\ToXYZr{1}{90}{45}) -- (top_of_plane);

            % Draw the angle between the dashed line and orange vector
            \pic[draw, <-, angle radius=4mm, angle eccentricity=1.2]{angle=velocity_vector_perpendicular--Sun--vertical_dashed_line};
            % Add a custom node for the label at the desired position
            \node at ($(Sun)!0.7!(velocity_vector_perpendicular)$) [xshift=-2mm, yshift=-1mm] {$\theta$};
        \end{scope}
    \end{tikzpicture}
    \caption{Orientation of a minihalo's initial velocity components. The Cartesian coordinate system is centered on the Galactic center ($\cal O$), with the Sun ($\star$) located along the positive $X$-axis and the Galactic disk in the $X-Y$ plane. The velocity vector \( \vec{v}_{\text{init}} \) is decomposed into a radial (\( \vec{v}_{\text{init}, \parallel} \)) component and a perpendicular (\( \vec{v}_{\text{init}, \perp} \)) component which lies on a plane perpendicular to the $X$-axis. The angle \( \theta \) determines the inclination of the perpendicular velocity relative to the $Z$-axis.}
    \label{fig:geometry}
\end{figure}
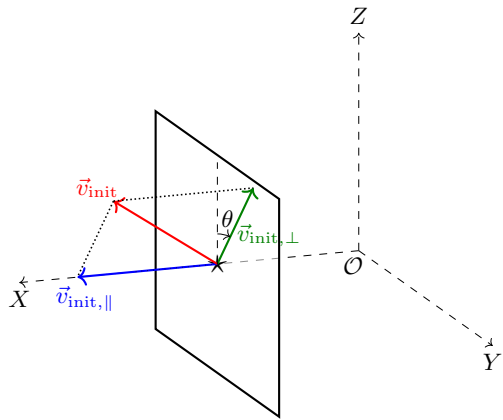

We still need two more pieces of information to fix the initial velocity vector. We first fix the galactic Cartesian coordinate system such that the Sun is located along the positive $X$-axis and the Galactic disk is in the $X-Y$ plane as shown in Fig.~\ref{fig:geometry}. Then, the vector corresponding to $v_{\mathrm{init,}\parallel}$ will be along the $X$-axis. Next, the vector corresponding to $v_{\mathrm{init,}\perp}$ lies in a plane parallel to the $Y$-$Z$ plane of the galactic coordinate system. Assuming this vector makes an angle $\theta$ with the positive $Z$-axis, we can decompose the vector along the $Z$-axis as $v_{\mathrm{init,}\perp} \cos(\theta)$ and along the $Y$-axis as $v_{\mathrm{init,}\perp} \sin(\theta)$. Now come two key steps. We choose the value of $\theta$ randomly in the interval $[0, 2\pi)$. The value of $\theta$ determines the inclination angle between the plane of orbit and the Galactic disk of the galaxy (which lies along the $X-Y$ plane). Furthermore, we choose the direction of the vector corresponding to $v_{\mathrm{init,}\parallel}$ randomly to be either along the positive or negative $X$-axis. With these two steps, we uniquely determine the orbit. Since our minihalo is randomly chosen to move clockwise or counter-clockwise, we are free to evolve our minihalo backward or forward in time. This is because in a gravitational system like the one we have here, there exists time-reversal symmetry. Evolving our minihalo backward in time with a given initial velocity vector is equivalent to evolving our minihalo forward in time with the initial velocity vector flipped in its direction, i.e., $t \to -t$ corresponds to $\vec{v}_{\mathrm{init}} \to -\vec{v}_{\mathrm{init}}$. In our code, we choose to evolve our minihalo forward in time using the singular isothermal potential given in Eq.~\ref{eq:singular isothermal sphere}. The details of how we did this are given in Appendix~\ref{app: Code to evolve an orbit}.

\section{Monte Carlo Simulations to determine the stellar-disrupted mass function of minihalos}

\label{sec: Monte Carlo Simulations to determine the stellar-disrupted mass function of minihalos}

In Section~\ref{sec: Formation halos from adiabatic perturbations}, we theoretically outlined how to determine the undisrupted mass function of minihalos. In this section, we follow a Monte-Carlo approach similar to that taken by \citelink{S2024}{S2024} to determine both the undisrupted mass function and the stellar-disrupted mass function of minihalos. However, here we take into account the potentially more destructive nature of multiple stellar encounters discussed in Sec.~\ref{sec: Accounting for multiple stellar encounters}. 

\subsection{Mass-Redshift Grid}\label{sec: mass-redshift grid}
We start off by creating a two-dimensional grid of virial mass ($M$) of minihalos and the infall redshift ($z$) at which the minihalos fell into their respective adiabatic halos. Thus, each grid point is associated with an ordered pair of $(M_i , z_j)$. We assign a randomly generated orbit to each grid point using the procedure in Section~\ref{sec: Monte Carlo sampling of orbits in singular isothermal sphere}. We choose the values of $M_i$ to be logarithmically spaced in the interval $[10^{-14}, 10^{-3}] M_\odot$. We choose the values of $z_j$ to be logarithmically spaced in $(1 + z)$ such that $z$ is in the range $[0, 150]$. We select 1,000 values of $M_i $ and 1,001 values of $z_j$ in those ranges.

 We approximate the integral in Eq.~(\ref{eq: undisrupted mass function}) by
\begin{equation}
	\label{eq: weight sum}
	\left. \frac{\mathrm{d} n_{\mathrm{f}}}{\mathrm{d} M}\right|_{M_i}=\sum_{j} w_{ij},
\end{equation}
where
\begin{equation}
	\label{eq: weights}
	w_{ij} = \delta z_{j} \frac{\mathrm{d} f_{\mathrm{adiab}}}{\mathrm{d} z}\left(z_{j}\right) \frac{\mathrm{d} n_0}{\mathrm{d} M}\left(M_i, z_{j}\right)\ ,
\end{equation}
and $\delta z_{j}=z_{j+1}-z_j$. Note that the value of $\delta z_{j}$ will vary with the $j$ considered since the $z$ values are logarithmically spaced. We will have one less number of $\delta z_{j}$ values than the number of $z_j$ values. Thus, $w_{ij}$ is evaluated using 1,000 values of $M_i$, and for each value of $M_i$, there will be 1,000 (not 1,001) values of $z_j$. Thus, we will have one million weights to evaluate.

\subsection{Energy Injection due to Stellar Interactions}

Each minihalo in our simulation will undergo some mass loss during its lifetime in our galaxy due to tidal interactions with the Galactic disk's stars. 
For a single pass through the Galactic disk, the time between individual stellar encounters will be much less than $t_{\rm dyn}$, so the energy injections can be added linearly, which means $p=2$ in Eq.~\ref{eq: Stucker et al's expression for effective energy injection parameter}.
\citelink{S2024}{S2024} estimate the injected energy for a passage through the disk to be
\begin{equation}\label{eq: E_frac for continuous stellar distribution}
	E_{\rm frac} =\frac{G m_\kappa \Sigma_*}{\sigma_*^2+v_{\mathrm{mh}}^2} \frac{\alpha^2(c)}{\gamma(c) \bar{\rho}_{\mathrm{vir}}(z)} \frac{2}{b_{\mathrm{s}}^2(c)+2 b_{\mathrm{C}}^2(\Sigma_*)}\ .
\end{equation}
where $G$ is the gravitational constant and 
\begin{align}
	b_{\rm s} = 6 \left(\frac{2\alpha(c)}{3\beta(c)}\right)^{1/2} r_{\rm vir}\, .
\end{align}
Also, 
\begin{align}
	\alpha&=\sqrt{\frac{c\left(-3-3 c / 2+c^2 / 2\right)+3(1+c) \ln (1+c)}{c^2(-c+(1+c) \ln (1+c))}}\\
    \beta &\approx \sqrt{\frac{c^2 \ln \left(r_{\mathrm{s}} / r_{\mathrm{c}}\right)+c^2 / 2-1 / 2}{\ln (1+c)-c /(1+c)}}\\
	\gamma&=\frac{c}{2} \frac{1-1 /(1+c)^2-2 \ln (1+c) /(1+c)}{[c /(1+c)-\ln (1+c)]^2}
\end{align}
where an NFW profile is assumed, and $r_{\mathrm{c}}$ is the smallest radius that the profile extends to, which is assummed to be $0.01 r_{\mathrm{s}}$.

As discussed in Section~\ref{sec: Mass-Concentration relationship}, given $M$ and $z$ one can work out the value of $c$.
The characteristic mass associated with the present-day mass function of stars in our Milky Way galaxy is $m_\kappa \sim 0.6 M_\odot$. The stellar surface density $\Sigma_*$ is obtained by integrating the stellar volume density $\rho_*$ along the one-dimensional trajectory of the minihalo within the galaxy as follows: 
\begin{equation}\label{eq: Sigma_* general formula}
	\Sigma_* = \int_\mathrm{traj} \rho_* \mathrm{d}l = \int_\mathrm{traj}\rho_* v \, \mathrm{d}t \ ,
\end{equation}
where ``traj'' represents the trajectory of the minihalo, $\mathrm{d}l$ represents the infinitesimal distance traveled by the minihalo along its trajectory, $v$ represents the instantaneous velocity magnitude of the minihalo relative to the Galactic disk, which we will assume to have static density distribution over the lifetime of the galaxy, and $\mathrm{d}t$ represents the infinitesimal time increment. The internal (to the galaxy) one-dimensional velocity dispersion of stars in the galaxy is denoted $\sigma_*$. The stars are assumed to have a Maxwell-Boltzmann velocity distribution function. Note that although the stars are moving in the galaxy, they do so as to keep the stellar volume density static on scales significantly larger than the average inter-stellar distance within the galaxy.  The variance of the velocity of the minihalo relative to the rest frame of the galaxy is denoted $v_{\rm mh}^2$. 
The term $\sigma_*^2+v_{\mathrm{mh}}^2$ has a weak dependence on the minihalo orbits and its value is assumed to be $(250\, \mathrm{km/s})^2$ in Eq.~(\ref{eq: E_frac for continuous stellar distribution}).

The shot noise is accounted for by the parameter
\begin{equation}\label{eq: b_c definition}
	b_{\rm C}\sim\sqrt{\frac{m_\kappa }{ \pi \Sigma_* }}\,.
\end{equation}

\subsection{Computing the Stellar Surface Density and Time-stamp of each Effective Disk Pass}

The Milky Way's disk consists of a thick disk and a thin disk. We model the stellar volume density in cylindrical coordinates with the Galactic center at the origin \cite{mcmillan2011mass}: 
\begin{equation}\label{eq: rho_* expression}
	\rho_*(R, Z)= \sum\limits_{d = {\rm thin, thick}} \frac{\Sigma_{\mathrm{d}, 0}}{2 Z_{\mathrm{d}}} \exp \left(-\frac{|Z|}{Z_{\mathrm{d}}}-\frac{R}{R_{\mathrm{d}}}\right) \ ,
\end{equation}
where $R, Z$ are the cylindrical coordinates of the galaxy.  The scale length is denoted $R_{\mathrm{d}}$, and $Z_{\mathrm{d}}$ is the scale height. They tell us how fast the stellar volume density falls off in the plane of the Galactic disk and perpendicular to the Galactic disk, respectively. The central surface density is denoted $\Sigma_{\mathrm{d}, 0}$. If we integrate out the $Z$-component of $\rho_*$, we will get the stellar surface density of the Milky Way as a function of $R$, and $\Sigma_{\mathrm{d}, 0}$ will be the stellar surface density at $R=0$. The parameters of Eq.~(\ref{eq: rho_* expression}) are given in Table~\ref{tab: paramters of rho_* expression}.
The $R$ in Eq.~(\ref{eq: rho_* expression}) is calculated as $R = \sqrt{X^2 + Y^2}$, where the $X$ and $Y$ axes define the plane of the Galactic disk.

\begin{table}
	\begin{tabular}{|l|c|c|}
		\hline
		&	Thin disk & Thick disk \\
		\hline
		$\Sigma_{\mathrm{d}, 0}\ [M_\odot/\mathrm{pc}^2]$ & 816.6 & 209.5 \\
		\hline
		$R_{\mathrm{d}}\ [\mathrm{kpc}]$ & 2.9 & 3.31\\
		\hline
		$Z_{\mathrm{d}}\ [\mathrm{kpc}]$ & 0.3 & 0.9\\
		\hline
	\end{tabular}
	\caption{
		\label{tab: paramters of rho_* expression}
		The parameters of Eq.(\ref{eq: rho_* expression}) are presented for both the thin and thick Galactic disks of the Milky Way galaxy. The parameter $\Sigma_{\mathrm{d}, 0}$ is the central surface density, $R_{\mathrm{d}}$ is the scale length, and $Z_{\mathrm{d}}$ is the scale height.
	}
\end{table}

\begin{figure}[htp]
	\centering
	\includegraphics[clip,width=\columnwidth]{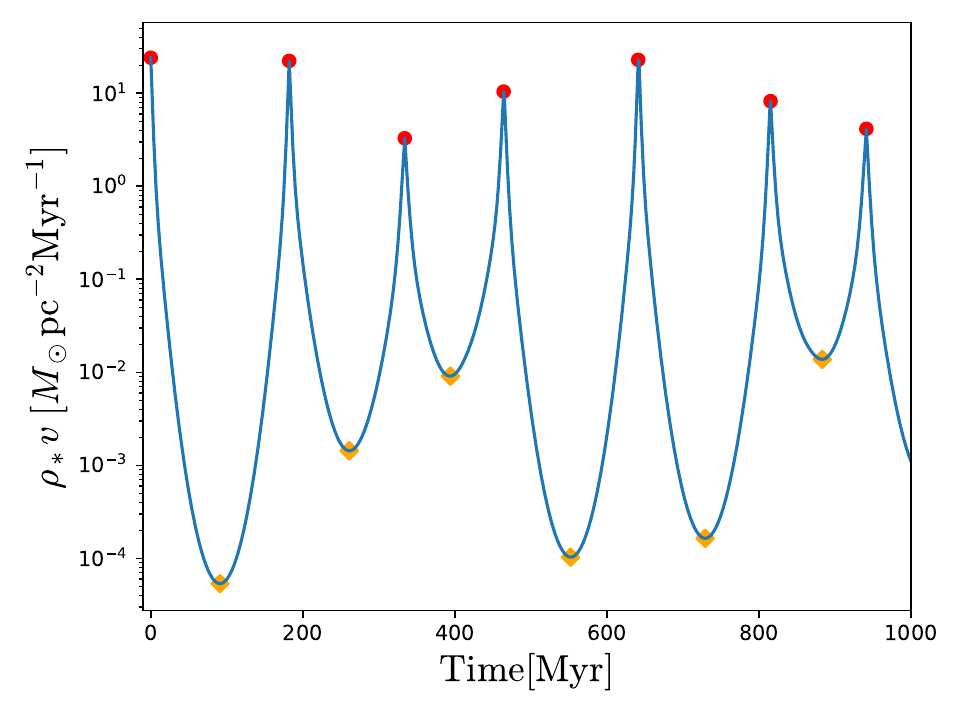}
	
	\hspace{0.5cm}\includegraphics[clip,width=\columnwidth]{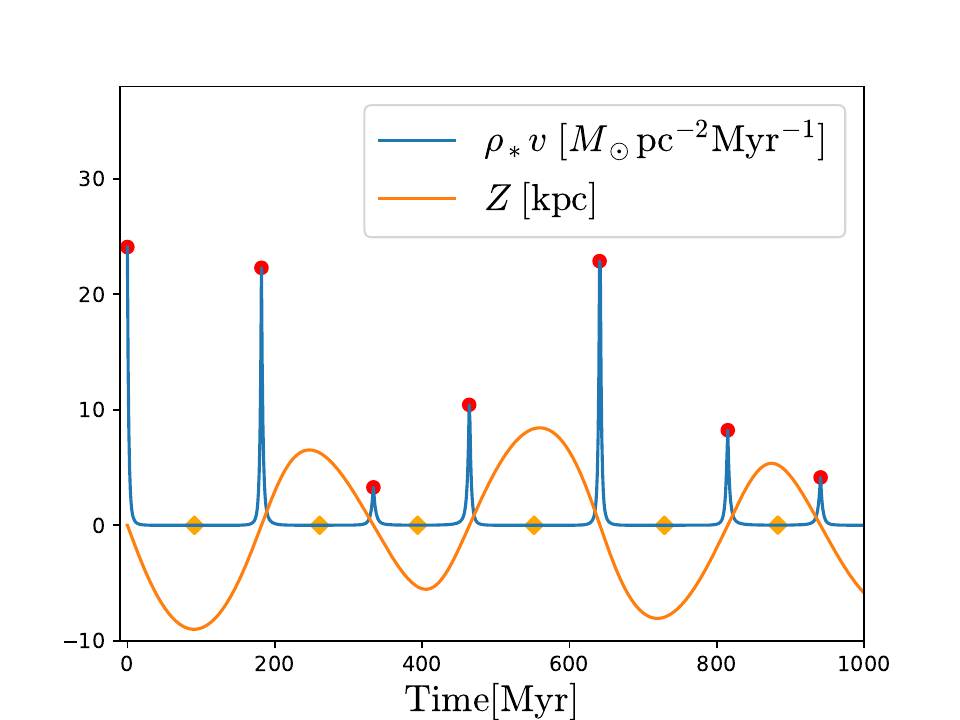}\hspace{-0.5cm}
	
	\caption{The minihalo is in orbit around the Galactic center. In the top panel, the integrand term $\rho_* v$ is plotted against time elapsed where $\rho_*$ is the stellar volume density at the position of the minihalo at any given time, and $v$ is the instantaneous speed of the minihalo at that time. The local maxima are denoted by a circular red dot. These points correspond to when the minihalo crosses the Galactic disk. The local minima are denoted by an orange diamond marker. In the bottom panel, additional information about the galactic $Z$ coordinate is provided.}
	\label{fig: integrand vs time - regular}
\end{figure}

Given the initial state parameters (position and velocity vectors) of a minihalo, we use the \texttt{lbparticles} code (see Appendix~\ref{app: Code to evolve an orbit}) to evaluate the position and velocity of the minihalo at certain discrete times in the future. Knowledge of the position vector at any instant of time allows us to evaluate the local stellar volume density $\rho_*$ at that time via Eq~(\ref{eq: rho_* expression}). Thus, at any given time, the integrand $\rho_* v$ (in Eq.~(\ref{eq: Sigma_* general formula})) corresponding to that time can be evaluated. The top panel of Fig.~\ref{fig: integrand vs time - regular} shows a sample minihalo in orbit around the Galactic center. The integrand $\rho_* v$ is plotted against time. We notice that the plot has local maxima and minima. The local maxima are denoted by a circular red marker, and in general correspond to the instant when a minihalo passes through the Galactic disk ($Z=0$). They are local maxima because the stellar volume density is highest at $Z=0$ for any given $R$ in Eq.~(\ref{eq: rho_* expression}). We consider a local maximum to be the time stamp of a single disk pass by the minihalo. On the other hand, the local minima in Fig.~\ref{fig: integrand vs time - regular} are denoted by an orange diamond marker, and in general correspond to being locally the furthest away from a disk pass that the minihalo can be at. The bottom panel of Fig.~\ref{fig: integrand vs time - regular} also plots the value of the galactocentric $Z$ coordinate versus time for that same orbit. It can be seen that, to a good approximation, the term $\rho_* v$ achieves a local maximum when $Z=0$ and a local minimum when $Z$ has a local extremum. We consider a single disk pass as being from one local minimum of the $\rho_* v$ curve to the next consecutive local minimum, with a local maximum in between. One thing to note is that the different local maxima do not have the same value of $\rho_* v$, which is evident from Fig.~\ref{fig: integrand vs time - regular}. This is because although the plane of the orbit remains fixed, the minihalo doesn't have closed orbits. Instead, the orbit precesses with time. This implies that the minihalo's multiple passes through the disk occur at different phases of the minihalo's orbit. Thus, the minihalo crosses the Galactic disk at different galactocentric radii $R$. Hence, the local stellar volume density $\rho_*$ at each disk pass will be different according to Eq.~(\ref{eq: rho_* expression}), creating different values for the integrand $\rho_* v$. Furthermore, it might be worth noting that the integrand $\rho_* v$ may not be a smooth function of time at the instant when $Z=0$. It is not smooth because $\rho_*$ is not smooth at $Z=0$ according to Eq.~(\ref{eq: rho_* expression}). Nonetheless, our procedure for finding the local maxima of $\rho_* v$ curve works because we use the discrete second difference in $\rho_* v$ values, and not the continuous second derivative.

Looking more closely at the bottom panel of Fig.~\ref{fig: integrand vs time - regular}, we see that the local maximum of the $\rho_* v$ curve is coincident with the instant of time when $Z=0$. This is because as $Z$ goes from negative to positive values through $Z=0$, $|Z|$ achieves a (non-smooth) local minimum at $Z=0$. Thus, from Eq.~(\ref{eq: rho_* expression}), we can see that the term $\exp\left(-|Z|\right)$ achieves a (non-smooth) local maximum at $Z=0$, hence forcing $\rho_*$ and consequently $\rho_* v$ to achieve a (non-smooth) local maximum here. On the other hand, the local minimum of the $\rho_* v$ curve is a little offset from the instant of time when the local extremum of the $Z$ curve occurs, i.e., $\dot{Z}=0$. This is because the integrand $\rho_* v$ is not just a function of $Z$ but also a function of $R$ and $v$ (see Appendix~\ref{app: local minimum of rho_*v does not coincide with z_dot=0} for details).

In our \texttt{Python} code using the \texttt{NumPy} library, we start with an array of linearly spaced time values with a resolution of approximately 1 Myr. We then evaluate the integrand $\rho_* v$ at each value of time. We then perform the following operation on the integrand array:  \texttt{diff}(\texttt{sign}(\texttt{diff}($\rho_* v$ array))). The \texttt{diff}() operation computes the difference between neighboring values of the input array. The \texttt{sign}() operation takes any real number as input and outputs $-1$ if the input is negative, $+1$ if the input is positive, and $0$ if the input is zero. The above net operation is a discrete form of the second derivative of the integrand $\rho_* v$ with respect to time. It is easy to see (with an example) that the output array of the net operation is two elements shorter than the original integrands array. Most of the values of this output array will be zero (corresponding to monotonic portions of the integrands vs time plot). If we find a value of $+2$ in this output array, the $\rho_* v$ curve achieves a local minimum at the position corresponding to that entry. On the other hand, if we find a value of $-2$ in the output array, the time instant corresponding to that entry is a local maximum. Thus, we are able to find the local extrema of the integrands array efficiently.

\subsection{Summing up the energy injection parameters for multiple Galactic disk passes}\label{sec: Summing up the energy injection parameters for multiple Galactic disk passes}

In Section~\ref{sec: mass-redshift grid}, we stated that in our Monte-Carlo simulations, we consider 1,000 values of $M$ and 1,000 values of $z$. Thus, we have one million ordered pairs or grid points of $(M, z)$. To each of these grid points, we assign one minihalo orbit that was generated using the procedure described in Section~\ref{sec: Monte Carlo sampling of orbits in singular isothermal sphere}. We also assign to each grid point a concentration parameter that depends on the $M$ and $z$ of that grid point according to Section~\ref{sec: Mass-Concentration relationship}. The goal is to calculate the effective energy injection parameter experienced by the orbit for each grid point during its lifetime in the galaxy. We then calculate the survival fraction of the minihalo given knowledge of this effective energy injection parameter and the concentration of the minihalo at $z$.

 Note that the time for which we calculate the effective energy injection parameter of the minihalo orbit corresponds to the lifetime of the orbit in the galaxy, which in turn corresponds to the infall redshift $z$. We must convert $z$ to its corresponding lookback time $T$ because the \texttt{lbparticles} code takes time (and not redshift) as input. We use the standard conversion formula (see, for example, Appendix~\ref{app: converting infall redshift to look back time}):
\begin{multline}\label{eq: look back time}
	T(z) = \frac{1}{H_0} \int_0^{z} \frac{1}{1+z^\prime} \\
	\times \left[\Omega_{\rm m}(1+z^\prime)^3 + \Omega_{\rm r}(1+z^\prime)^4 + \Omega_\Lambda\right]^{-1/2} \mathrm{d}z^\prime
\end{multline}

We numerically integrate Eq.~(\ref{eq: look back time}) using the \texttt{SciPy}'s \texttt{quad}() function in Python. We now can use the \texttt{lbparticles} code to evolve the orbit of the minihalo from the infall redshift until today.

\section{Computing the survival fraction of minihalos}\label{sec: computing the survival fraction of the minihalo}

In \citelink{dsouza2024}{Paper 1}, we presented a procedure to evaluate $\Delta M/M$, for an NFW minihalo, given $c$ and $E_{\rm frac}$
where $\Delta M$ is the mass loss due to the energy injection.
However, the ranges of concentration and $E_{\rm frac}$ values considered in that article were rather narrow compared to what is generated by the Monte-Carlo simulation in this article. Thus, we generate an interpolation object offline using Python's \texttt{RegularGridInterpolator} function from the \texttt{SciPy.Interpolate} package that takes as input the energy injection parameter and concentration and outputs the survival fraction. To achieve this, we first create a log-spaced array of concentration values in the range $c \in [0.1, 10^5]$ and similarly for energy injection parameters values in the range $E_{\rm frac} \in [10^{-13}, 3\times10^6]$. We then use \texttt{NumPy}'s \texttt{meshgrid} function to generate all possible ordered pairs of $(E_{\rm frac}, c)$. Then, we use the procedure in \citelink{dsouza2024}{Paper 1} to compute the survival fraction corresponding to each ordered pair. This can be a time-consuming process, but it is done offline and only once. However, once the interpolation object is generated, it can compute the survival fraction rapidly as long as the input $E_{\rm frac}$ and $c$ are in the ranges that were originally used to generate the interpolation object in the first place.

We can integrate the mass function, $\frac{\mathrm{d}n_{f}}{\mathrm{d}M}$, to get the collapsed fraction $f$ which is the ratio of the number of axions that have collapsed into minihalos to the total number of axions. 
As done by \citelink{S2024}{S2024}, we choose the lower bound of this mass range to be $10^{-12}M_\odot$. 
Our conclusions about the ratio of the collapsed fraction with and without disruption will not be sensitive to making this lower bound even lower.
The upper bound of the mass range is chosen to be the same as the upper bound that we used for the Monte-Carlo simulation, i.e., $10^{-3}M_\odot$. This value is motivated by wanting our most massive minihalo to be substantially smaller than our least massive adiabatic halo. Therefore, the original collapsed fraction without stellar disruption is given by 
\begin{equation}\label{eq: expression for 'f'}
	f_{\rm ori} = \frac{1}{\bar{\rho}_{\rm c}}\int_{10^{-12}M_\odot}^{10^{-3}M_\odot} M \frac{\mathrm{d}n_{f}}{\mathrm{d}M}\ \mathrm{d}M \, .
\end{equation}
where $\bar{\rho}_{\rm c}$ is the comoving density of cold dark matter. 
% Substituting Eq.~\ref{eq: undisrupted mass function} into the above equation gives
% \begin{equation}\label{eq: expression for 'fundisrupted'a}
% 	\begin{aligned}
% 		f_{\rm ori} =& \frac{1}{\bar{\rho}_{\rm c}}\int_{10^{-12}M_\odot}^{10^{-3}M_\odot}\int_{z_{\mathrm{eq}}}^0  M \frac{\mathrm{d} f_{\rm adiab}}{\mathrm{d} z}\left(z\right) \times \\&        
% 		  \frac{\mathrm{d} n_0}{\mathrm{d} M}\left(M, z\right)\, {\rm d}  M \mathrm{d} z\,.    
% 	\end{aligned}
% \end{equation}
Using Eqs.~(\ref{eq: weight sum}) and (\ref{eq: weights}), the above equation can be approximated by
\begin{align}\label{eq: expression for 'fundisrupted'b}
	f_{\rm ori} &= \frac{1}{\bar{\rho}_{\rm c}} \sum_{i,j} M_i          w_{ij} \delta M_i\nonumber \\
	&= \frac{1}{\bar{\rho}_{\rm c}}\sum_{i,j}       M_i^2   w_{ij} \delta \ln(M),
\end{align}
where $\delta M_i = M_{i+1} - M_i$ is the mass bin width and $\delta\ln(M)\approx\delta M_i/M_i$ is the constant bin width for $\ln(M)$.
We can then approximate the derivative with respect to the  mass as
\begin{equation}\label{eq: expression for 'dfundisrupt'}
	\begin{aligned}
		\left.\frac{{\rm d}f_{\rm ori}}{{\rm d\,ln}(M)}\right|_{M_k} &= \frac{1}{\bar{\rho}_{\rm c}} M_k^2\sum_j           w_{kj}\, .
	\end{aligned}
\end{equation}
For plotting purposes we use 
\begin{equation}\label{eq: mass function definition, df/dM - part 1}
	\frac{\mathrm{d}f}{\mathrm{d}\log_{10}(M)} = \ln(10) \frac{\mathrm{d}f}{\mathrm{d}\ln (M)} %
	\ .
\end{equation}

\begin{figure}
	\includegraphics[width=\columnwidth]{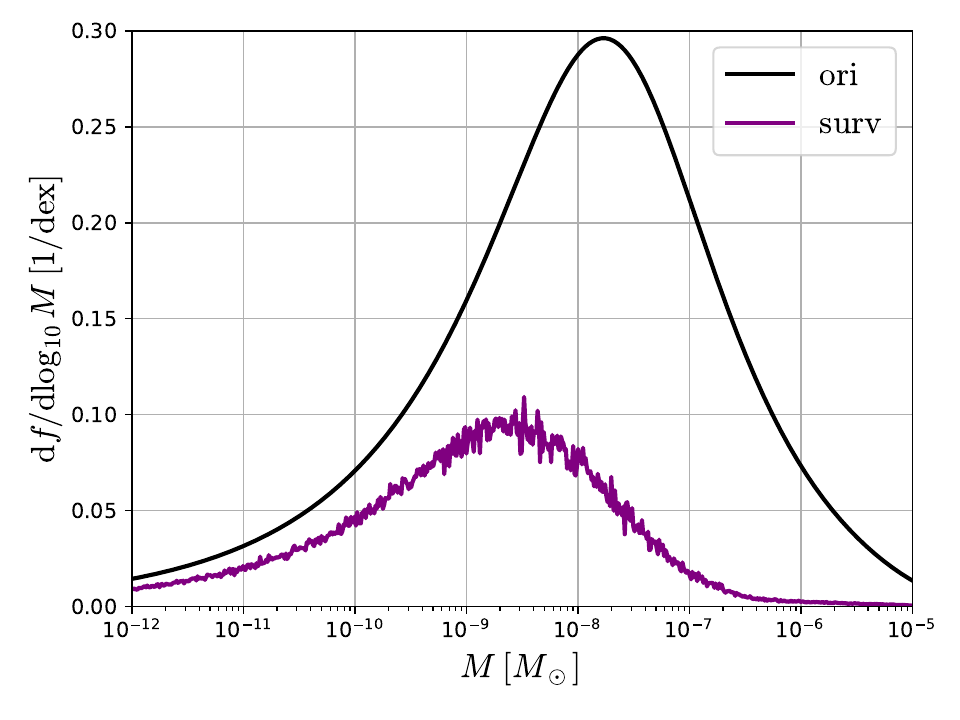}
\caption{The mass function for $m_a=25$~$\mu$eV axion minihalos of mass  $M$ and collapse fraction $f$. In this figure, we have assumed the minimum adiabatic halo mass is $10^2M_\odot$. The mass function without considering any disruption is shown as a black curve, while the mass function incorporating the effects of stellar disruption is shown as a purple curve. In generating the disrupted mass functions, we used the hybrid method to combine multiple energy injections.  }
	\label{fig: mass_function}
\end{figure}

Combining Eqs.~(\ref{eq: mass function definition, df/dM - part 1}) and (\ref{eq: expression for 'dfundisrupt'}), we get: 
\begin{equation}\label{eq: S2024 mass function in terms of ordinary mass function}
	\left. \frac{\mathrm{d}f_{\rm ori}}{\mathrm{d}\log_{10}(M)}\right|_{M_k} =  \frac{\ln(10)}{\bar{\rho}_{\rm c}} M_k^2   \sum_j        w_{kj}\,  .
\end{equation}
This quantity is plotted in Fig.~\ref{fig: mass_function} as the black curve. It is a good match to the analogous dashed gray curve in the top panel of Fig.~10 of \citelink{S2024}{2024}. 

To generate the mass function in the presence of stellar disruption, we first compute $\Delta M_{ij}$ for the minihalo orbit corresponding to each grid point.

Similarly to Eq.~(\ref{eq: expression for 'fundisrupted'b}) the surviving disrupted fraction of axions in minihalos is given by
\begin{align}\label{eq: expression for 'f'b}
	f_{\rm surv} =& \frac{1}{\bar{\rho}_{\rm c}} \sum_{i,j} (M_i - \Delta M_{ij})          w_{ij}\delta M_i\nonumber \\
	=& \frac{1}{\bar{\rho}_{\rm c}} \sum_{i,j} (M_i - \Delta M_{ij})          w_{ij}M_i\delta\ln(M)\nonumber \\
	=& \frac{\delta\ln(M)}{\bar{\rho}_{\rm c}} \times \nonumber\\
	&\sum_k \sum_{M_k\leq M_i-\Delta M_{ij}<M_{k+1}} (M_i - \Delta M_{ij})   M_i       w_{ij},
\end{align}
where the sum in the last line of the above equation is over all $i,j$ that satisfy $M_k\leq M_i-\Delta M_{ij}<M_{k+1}$.
We can then approximate the derivative with respect to the mass as
\begin{equation}\label{eq: expression for 'df'}
	\begin{aligned}
		\left.\frac{{\rm d}f_{\rm surv}}{{\rm d\,ln}(M)}\right|_{M_k} =& \frac{1}{\bar{\rho}_{\rm c}} \times \\
		&\sum_{M_k\leq M_i-\Delta M_{ij}<M_{k+1}} (M_i - \Delta M_{ij})M_i          w_{ij}\, .
	\end{aligned}
\end{equation}
Using (\ref{eq: mass function definition, df/dM - part 1}) with the above equation gives
\begin{equation}
	\begin{aligned}
		\label{eq: mass fraction step by step - part 3}
		\left.\frac{{\rm d}f_{\rm surv}}{{\rm d\,}\log_{10}(M)}\right|_{M_k} =& \frac{\ln(10)}{\bar{\rho}_{\rm c}}\times \\
		&\sum_{M_k\leq M_i-\Delta M_{ij}<M_{k+1}} (M_i - \Delta M_{ij})M_i          w_{ij}\, .
	\end{aligned}
\end{equation}

Eq.~(\ref{eq: mass fraction step by step - part 3})  gives us the alternative form of the stellar disrupted mass function of minihalos in the galaxy today. This mass function is presented in Fig.~\ref{fig: mass_function} as the purple curve. The stellar-disrupted mass function is significantly more suppressed than the undisrupted mass function. Moreover, the peak in the stellar-disrupted mass function has shifted towards lower mass relative to the undisrupted mass function. The reason for both these observations is that the minihalos lose mass when subjected to stellar interactions. Finally, the reason that the stellar-disrupted curve is noisy is that we have assigned a random minihalo orbit to each $(M, z)$ grid point. If the Monte-Carlo simulation is run multiple times independently, the exact values taken by this curve change but only up to the statistical noise induced by the noise in the energy and angular momentum of the generated orbits. As we will see later in the section, our final results will be insensitive to this noise.

The disrupted (purple) curve in Fig.~~\ref{fig: mass_function} can be compared to the red curve in Fig.~10 of \citelink{S2024}{S2024}. They also considered a smooth tidal disruption from the Milky Way potential. But as can be seen from the same figure, including that had a negligible effect once the stellar tidal effects were accounted for. Their disrupted curve has \textcolor{black}{about double}  the area of our one as they added the energy injections linearly, equivalent to our $p=2$ case. Their disrupted curve is also smoother than ours. The reason for this is that they used correction factors based on Monte Carlo averages. We couldn't employ that technique as it was incompatible with our hybrid method of adding multiple energy injections.

% \section{Computing the surviving mass in minihalos in the galaxy}\label{sec: computing the surviving mass in minihalos in the galaxy}

To get a measure of how much disruption has taken place, we evaluate 
\begin{equation}
	\frac{M_{\rm surv}}{M_{\rm ori}} = \frac{f_{\rm surv}}{f_{\rm ori}} 
\end{equation}
where $M_{\rm surv}$ is the amount of mass in minihalos 8~kpc from the Galactic center today, and $M_{\rm ori}$ would be the amount of mass in minihalos 8~kpc from the Galactic center today if stellar disruption had not taken place.

Using the minimum mass of adiabatic halos considered to be $10^{2} M_\odot$ for the Monte-Carlo simulation, we find that $M_{\rm surv} / M_{\rm ori} = 30\%$ for $m_a=25\mu eV$. On the other hand, \citelink{S2024}{S2024} states that this value is $58\%$ using their method of correction factors and setting $p=2$.  This lost mass enters the inter-minihalo space called minivoids. Thus, the axion density in the minivoids increases. Despite the stellar disrupted mass function in Fig.~\ref{fig: mass_function} being noisy and its exact values changing slightly in between independent runs of the Monte-Carlo simulation, the resulting value of $M_{\rm surv} / M_{\rm ori}$ does not change in between runs at the level of $0.1\%$.

It is expected that the average inter-minihalo distance is significantly greater than the virial radius of a typical minihalo. Thus, the Earth is more likely to be inside a minivoid than inside a minihalo (e.g., \cite{Eggemeier23,ohareAxionMiniclusterStreams2024}). Thus, the lower value of $M_{\rm surv} / M_{\rm ori}$ means that the local axion density at the Earth's position is likely to be higher than previously predicted. This increases the chances of axion dark matter direct detection using haloscopes relative to what had been previously estimated without considering the more destructive nature of multiple stellar encounters.

\begin{figure}[htp]
	\centering
	\includegraphics[clip,width=\columnwidth]{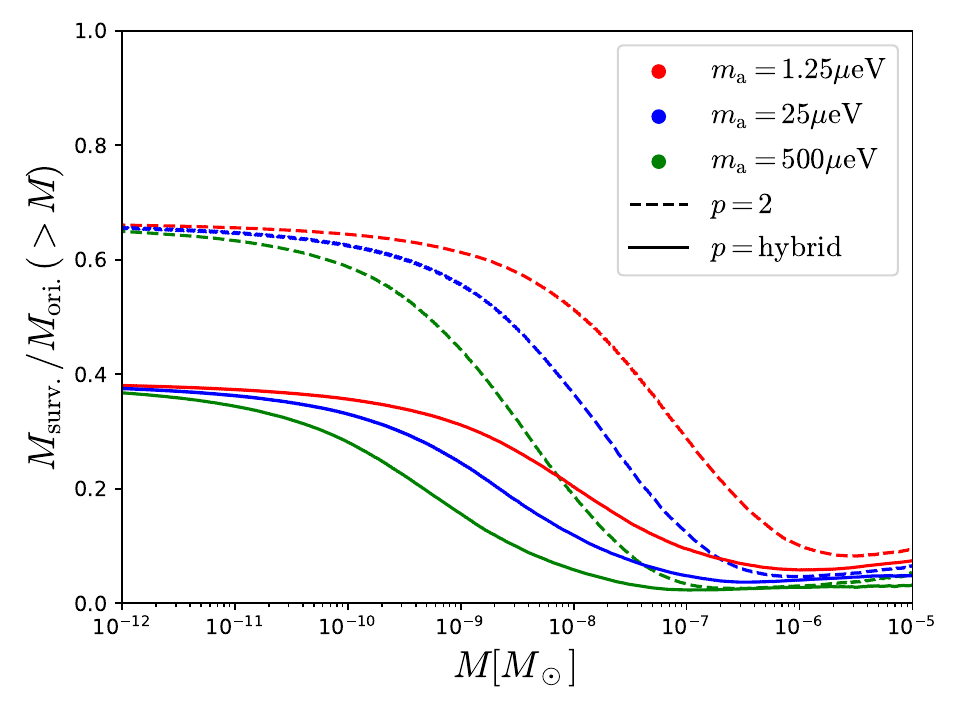}
	
	\hspace{0.5cm}\includegraphics[clip,width=\columnwidth]{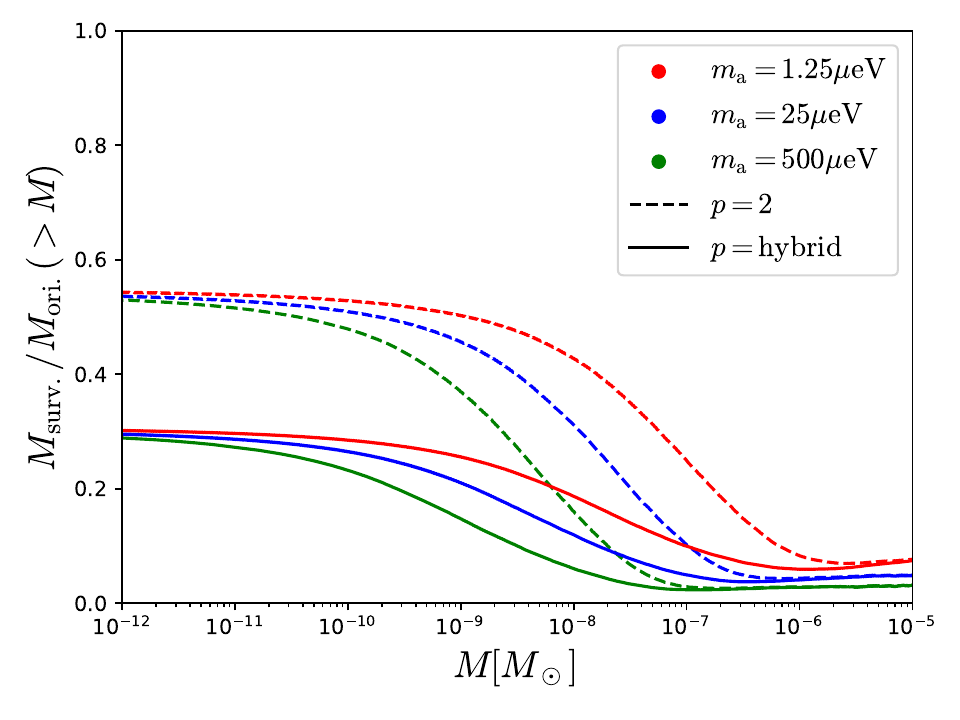}
	
	\caption{The value of $M_{\rm surv} / M_{\rm ori} (>M)$ is plotted against the minimum minihalo mass $M$. The parameter $M_{\rm ori}$ would be the mass at $8$ kpc of minihalos if there was no stellar or other tidal  disruption. While $M_{\rm surv}$ is the corresponding mass if stellar disruption is accounted for. The top and bottom panels use the lower bound in the mass of adiabatic halos to be equal to $10^{-2}$ and $10^2 M_\odot$ respectively. The results for three different axion masses $m_{\rm a}=1.25, 25, 500 \mu \text{eV}$ are presented as different colors. The dashed lines represent the $p=2$ method of linearly adding up the energy injection parameters corresponding to consecutive disk passes by the minihalo. The solid lines represent the ``hybrid" method of adding up the energy injection parameters.}
	\label{fig: M_surv by M_ori vs M}
\end{figure}

In Fig.~\ref{fig: M_surv by M_ori vs M}, we have calculated the values of $M_{\rm surv} / M_{\rm ori}$ 
as a function of the lower bound of the 
integration in Eq.~(\ref{eq: expression for 'f'}). 
%The top/bottom panel uses the lower bound $M_{\rm min}$ on the mass of adiabatic halos to be equal to $10^{-2} / 10^2 M_\odot$. In each panel, the different colors represent different axion masses $m_{\rm a}$. We have considered $m_{\rm a} = 1.25, 25, 500 \mu \text{eV}$. The dashed lines use $p=2$, which amounts to linearly adding up the energy injection parameters for consecutive disk passes for a given minihalo orbit in the Monte-Carlo simulation. We present this case because \citelink{S2024}{S2024} effectively uses $p=2$ although they do not evolve each minihalo orbit. On the other hand, the solid lines represent our method of either linearly or non-linearly adding up the energy injection parameters of consecutive disk passes by comparing the time between consecutive disk passes to the dynamical time of the minihalo. We call this method the $p=\text{``hybrid"}$ method. From Fig.~\ref{fig: M_surv by M_ori vs M}, 
We see from this figure that 
the $p=\text{``hybrid"}$ method results in lesser mass surviving in minihalos compared to the $p=2$ method. We also see from the figure that when we set that adiabatic minimum halo mass $M_{\rm min} = 10^2 M_\odot$, we find that lesser mass in minihalos survives compared to setting $M_{\rm min} = 10^{-2}M_\odot$. This is because, as can be seen from Fig.~\ref{fig: collapse fraction}, the $M_{\rm min}=10^2 M_{\odot}$ case will freeze at a lower redshift.
From Fig.~1 of \citelink{S20204}{S2024}, we can see that this implies the concentration of the minihalos in the $M_{\rm min}=10^2M_{\odot}$ case will be higher. This is confirmed in Fig.~\ref{fig: contour plot of joint PDf of Efrac and c} in Appendix~\ref{app: Distribution of energy injection parameters and concentrations}
where we plot the 95\% contours of the joint distribution of $E_{\rm frac}$ and $c$.
As can be seen from this figure, most of the minihalos will have an effective $E_{\rm frac}<1$. As shown from Fig.~4 of \citelink{S2024}{S2024}, the higher concentration minihalos will suffer a greater mass loss. This, and the difference in $E_{\rm frac}$, explains why the $M_{\rm min}=10^2M_{\odot}$ case is found to have a greater mass loss in minihalos in comparison to the $M_{\rm min}=10^{-2}M_\odot$ case. 

We see that as we increase the mass $M$ in Fig.~\ref{fig: M_surv by M_ori vs M} from $M=10^{-12}M_\odot$, the value of $M_{\rm surv} / M_{\rm ori}$ decreases. The reason for this trend becomes clearer when we look at Fig.~\ref{fig: mass_function}, where the undisrupted mass function rises more sharply relative to the stellar-disrupted mass function. This causes $M_{\rm surv} / M_{\rm ori}$ to decrease. However, in Fig.~\ref{fig: M_surv by M_ori vs M}, when we go to high masses of the order of $\gtrsim 10^{\textcolor{black}{-}7} M_\odot$, we see that the value of $M_{\rm surv} / M_{\rm ori}$ starts to increase. This is because in Fig.~\ref{fig: mass_function}, the value of the stellar-disrupted mass function begins to level off around $M \gtrsim 10^{-7} M_\odot$ while the undisrupted mass function still keeps dropping. This causes the value of $M_{\rm surv} / M_{\rm ori}$ to increase.

\begin{figure}
	\includegraphics[width=\columnwidth]{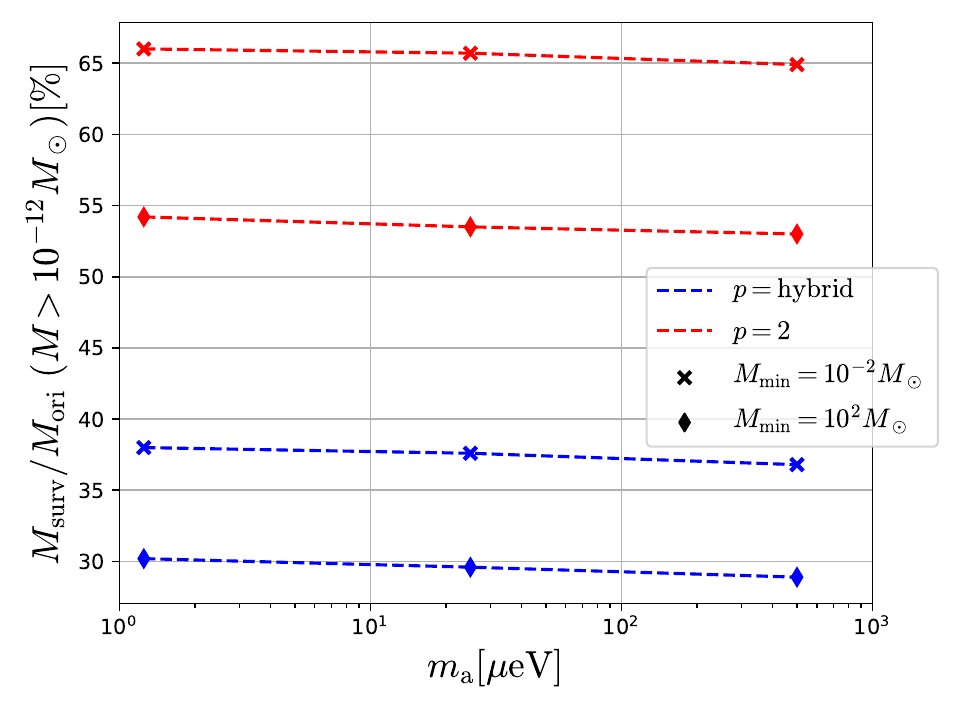}
	\caption{
    %Information similar to Fig.~\ref{fig: M_surv by M_ori vs M} is presented, except that the lower bound in the integral in Eq.~(\ref{eq: expression for 'f'}) is set to $10^{-12}M_\odot$ and not $M$. $M_{\rm surv} / M_{\rm ori}
    A plot of $M_{\rm surv} / M_{\rm ori}(M>10^{-12}M_\odot)$ against discrete values of the axion mass $m_{\rm a}=1.25, 25, 500 \mu \text{eV}$. The cross and diamond markers represent the case where the lower bound $M_{\rm min}$ on the adiabatic halos is $10^{-2}$ and $10^2 M_\odot$, respectively. The red and blue lines represent the $p=2$ and hybrid method, respectively, of summing up the energy injection parameters of consecutive Galactic disk passes.}
	\label{fig: M_surv by M_ori greater than 1e-12 MSolar vs axion mass}
\end{figure}

The dashed curves in the bottom panel of Fig.~\ref{fig: M_surv by M_ori vs M} show reasonable agreement with the corresponding orange curves in Fig.~14 of \citelink{S2024}{S2024}. The minor differences are likely due to methodological distinctions: \citelink{S2024}{S2024} generated probability density functions (PDFs) from their Monte Carlo simulations and subsequently derived correction factors from those PDFs, whereas we directly implemented the Monte Carlo simulations without relying on intermediate corrections.

Next, we combine both the panels of Fig.~\ref{fig: M_surv by M_ori vs M} and only present the value of $M_{\rm surv} / M_{\rm ori}$ where the lower bound of the integral in Eq.~(\ref{eq: expression for 'f'}) is set back to $10^{-12}M_\odot$. This is presented in Fig.~\ref{fig: M_surv by M_ori greater than 1e-12 MSolar vs axion mass} where we plot $M_{\rm surv} / M_{\rm ori} (M>10^{-12}M_\odot)$ against axion mass $m_{\rm a}$ for discrete axion masses $m_{\rm a} = 1.25, 25, 500 \mu \text{eV}$. 
%The cross/diamond markers represent the case of lower bound on the mass of adiabatic halos $M_{\rm min} = 10^{-2}/10^2 M_\odot$. The blue lines represent the case of $p=\text{``hybrid"}$ and the red lines represent the case of $p=2$.
This information is also presented in Table~\ref{table:mass ratios}. We checked the Monte-Carlo simulation convergence
was insensitive to small changes in 
the maximum mass of the minihalos, maximum infall redshift, and number of sample points of mass and infall redshifts. \textcolor{black}{Finally, note that in reality, we expect the dynamical time of a minihalo to increase slightly after each disk pass due to a change in the density profile of the minihalo. To be computationally efficient, we have not considered this effect. However, in Appendix~\ref{app:updating dynamical time}, using a back-of-the-envelope calculation, we have estimated that the true value of $M_{\rm surv}/M_{\rm ori}$ will not increase by more than a few percent compared to the values presented in Table~\ref{table:mass ratios} if we did update the dynamical time.}

\begin{table}
	\begin{tabular}{|l|c|l|l|}
		\hline
		$M_{\rm min} [M_\odot]$&	p & $m_{\rm a} [\mu \rm eV]$ & $M_{\rm surv} / M_{\rm ori}(M>10^{-12}M_\odot)[\%]$  \\
		\hline
		$10^{-2}$ &  2 & 1.25  & 66  \\
		\hline
		$10^{-2}$ &  2 &  25 & 65.7  \\
		\hline
		$10^{-2}$ &  2 &  500 &  64.9 \\
		\hline
		$10^{-2}$ & hybrid  & 1.25  &  38 \\
		\hline
		$10^{-2}$ & hybrid &  25 &  37.6 \\
		\hline
		$10^{-2}$ &  hybrid & 500  &  36.8 \\
		\hline
		$10^2$ & 2  & 1.25  &  54.2 \\
		\hline
		$10^2$ & 2  &  25 &  53.5 \\
		\hline
		$10^2$ & 2  &  500 &  53 \\
		\hline
		$10^2$ & hybrid  &  1.25 &  30.2 \\
		\hline
		$10^2$ & hybrid  &  25 &  29.6 \\
		\hline
		$10^2$ & hybrid  & 500  & 28.9  \\
		\hline
	\end{tabular}
	\caption{
    A table of $M_{\rm surv} / M_{\rm ori}(M>10^{-12}M_\odot)$. 
    See Fig.~\ref{fig: M_surv by M_ori greater than 1e-12 MSolar vs axion mass} for a plot of these values.
		\label{table:mass ratios}
	}
\end{table}

\section{Conclusions}
\label{sec: conclusions}

In this work, we have investigated the disruption of axion minihalos due to stellar encounters in the Milky Way galaxy. We extended previous analyses by incorporating a more accurate treatment of multiple stellar encounters, taking into account whether minihalos have sufficient time to relax between encounters based on their dynamical timescales. By generating a population of minihalo orbits using Monte Carlo simulations and evolving them within a model of the Galactic potential, we computed the stellar-disrupted mass function of minihalos.

Our results indicate that the cumulative effect of stellar interactions is more destructive to minihalos than previously estimated. Specifically, we find that the surviving mass fraction $M_{\rm surv} / M_{\rm ori}$ of minihalos is significantly reduced when accounting for the proper addition of energy injections from multiple stellar encounters. For example, when using a minimum mass of adiabatic halos of $M_{\rm min} = 10^2\, M_\odot$, we find that only about $30\%$ of the original mass in minihalos survives, compared to previous estimates of around $60\%$. This reduction is due to the increased mass loss when minihalos have time to relax between encounters, leading to a nonlinear addition of energy injections.

The suppression of the stellar-disrupted mass function, as illustrated in Fig.~\ref{fig: mass_function}, has important implications for the distribution of axion dark matter in the Galaxy. With a larger fraction of axion dark matter residing in the inter-minihalo space (minivoids), the local axion density at the Earth's position may be higher than previously predicted. This enhancement increases the prospects for the detection of axions via haloscopes relative to the case where the more destructive nature of multiple stellar encounters was not accounted for.

Furthermore, our analysis highlights the importance of accurately modeling the cumulative effects of stellar encounters on minihalos. By considering the dynamical timescales of minihalos and adopting a hybrid method for summing energy injections (as opposed to the linear addition method with $p=2$), we provide a more realistic estimate of minihalo survival.

Future work could extend this study by incorporating the streaming effect of the minihalo disruption as was done, for example, in Ref.~\cite{ohareAxionMiniclusterStreams2024}. However, there the more destructive effects of multiple encounters were not accounted for.

In conclusion, our findings suggest that stellar disruption plays a significant role in shaping the minihalo mass function and the distribution of axion dark matter in the Milky Way. Accurately accounting for these effects is crucial for interpreting observational data and guiding the search for dark matter.

\begin{acknowledgments}
	ID is supported by a University of Canterbury Doctoral Scholarship. JCF is grateful for support from the New Zealand Government, administered by the Royal Society Te Ap\={a}rangi. 
\end{acknowledgments}

\section*{Data Availability}
The data and code that support the findings of this article are openly available \cite{dsouza_2025_zenodo}.

\bibliography{axion_minihalo.bib}

\appendix

\section{Rescaling the \texttt{CAMB} isocurvature growth function}\label{app: renormalizing the CAMB growth function}
\citelink{X2021}{X2021} gave the following approximate formula for the isocurvature growth function: 
\begin{equation}\label{eq: linear growth function}
	D(a) = \frac{2}{3} + \frac{a}{a_{\rm eq}}\ ,
\end{equation}
where $a$ is the scale factor at which the growth function is evaluated, and $a_{\rm eq}$ is the scale factor at matter-radiation equality. This growth function varies linearly with the scale factor $a$. It does not take into account the contribution from dark energy and hence is not accurate during the dark energy-dominated epoch (and consequently today). Dark energy causes the growth function to become sub-linear at late times. The isocurvature growth function that we calculate using \texttt{CAMB} is accurate at all redshifts in consideration. 
%Therefore, we rescale the \texttt{CAMB} growth function to match the linear growth function in Eq.~(\ref{eq: linear growth function}) in the matter-dominated epoch, thus generating a \textit{rescaled CAMB} growth function which can be used with \citelink{X2021}{X2021}'s formulas for the pre-infall mass function that are given in the next section.

\begin{figure}
	\includegraphics[width=\columnwidth]{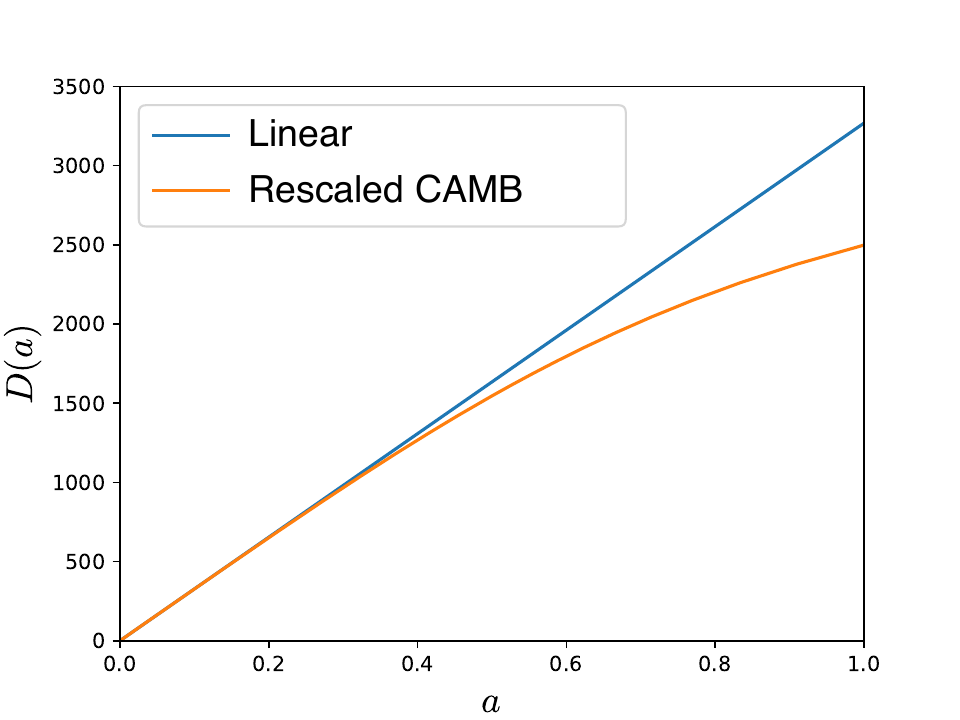}
	\caption{The isocurvature growth function $D$ is shown as a function of the scale factor $a$. The linear growth function is a solution to the M\'esz\'aros equations and doesn't take into account the dark energy contribution. The rescaled \texttt{CAMB} growth function is obtained by using the \texttt{CAMB} package. It takes into account the dark energy contribution. The two growth functions match during the matter- and radiation-dominated epochs but differ in the dark energy-dominated epoch.}
	\label{fig: growth_factor}
\end{figure}

The \texttt{CAMB} growth function,  $D_{\rm CAMB}$ is calculated using the power spectrum at some reference redshift (in our case, $z=100$) in the matter-dominated epoch.
For $z\gg 1$,
\begin{equation}
    D_{\rm CAMB}(z)=AD(z)
    \label{eq:DCAMB}
\end{equation}
where $A$ is some normalization constant to be determined and $D(a)$ is given in Eq.~(\ref{eq: linear growth function})).
We calculate $A$ using two reference redshifts:  $z=10$ and $z_{\rm eq} = 3266$ (the redshift of matter-radiation equality). Thus, using Eqs.~\ref{eq: linear growth function} and \ref{eq:DCAMB} we get
\begin{equation}
	 A=a_{\rm eq}\frac{D_{\rm CAMB}(a(z=10)) - D_{\rm CAMB}(a_{\rm eq})}{a(z=10) - a_{\rm eq}} \ ,
 \label{eq:evaluating A}    
\end{equation}
where $a(z) = 1 / (1+z)$, and $a_{\rm eq} = a(z_{\rm eq})$.  

Let $\widetilde{D}_{\rm CAMB}$ be the rescaled CAMB growth function:
\begin{equation}\label{eq: preliminary expression for renormalized growth function}
	\widetilde{D}_{\rm CAMB}(a) = \frac{{D}_{\rm CAMB}(a)}{ A}\ ,
\end{equation}
where $A$ is obtained from Eq.~\ref{eq:evaluating A}.

Fig.~\ref{fig: growth_factor} shows the growth function $D$ as a function of the scale factor $a$. It can be seen that the rescaled \texttt{CAMB} growth function is linear during the matter and radiation-dominated epochs but becomes suppressed to sub-linear during the dark energy-dominated epoch.

\section{Generating the pre-infall mass function}\label{app: pre-infall mass function}
Here, we generate the pre-infall mass function of axion minihalos using a modified Sheth-Tormen formalism \citelink{X2021}{X2021}. For minihalos that do not get captured by the halo of a galaxy, the  mass function $\mathrm{d}n_0/\mathrm{d}M$ is governed by the following equation: 
\begin{equation}\label{eq: mass_function_governing_equation}
	\frac{M^2 (\mathrm{d} n_0 / \mathrm{d} M)}{\bar{\rho}_{\rm c}} \frac{\mathrm{d} M}{M}=\nu f(\nu) \frac{\mathrm{d} \nu}{\nu} \, ,
\end{equation}
where $M$ represents the mass of a minihalo and $\bar{\rho}_{\rm c}$ is the comoving density of cold dark matter.
The preinfall comoving number density of minhalos between masses $M_{\rm min}$ and $M_{\rm max}$ is given by
\begin{equation}
     \int_{M_{\rm min}}^{M_{\rm max}} {\rm d}M\, \frac{\mathrm{d} n_0}{\mathrm{d}M} \,.
\end{equation}
 The parameter $\nu$ is defined by: 
\begin{equation}\label{eq: nu definition}
	\nu(M, z) \equiv \frac{\delta_{\rm c}^2}{\sigma^2(M, z)} \ ,
\end{equation}
where $\delta_{\rm c} = 1.686$ is the critical overdensity for spherical collapse of axion density perturbations and  $\sigma^2(M, z)$ is the variance of the initial density perturbations when smoothed with a spherical top-hat filter of length scale $R=(3 M / 4 \pi \bar{\rho}_{\rm c})^{1 / 3}$. It can be shown that the variance of the primordial white-noise matter power spectrum corresponding to the axion is given by \citelink{X2021}{X2021} and  \citelink{S2024}{S2024} as 
\begin{equation}\label{eq: sigma definition}
	\sigma(M, z)=D(z) \sqrt{\frac{3 A_{\mathrm{osc}}}{2 \pi^2} \frac{M_0}{M}} \ ,
\end{equation}
where $D$ is the isocurvature growth function of axion density perturbations. The amplitude of the white-noise matter power spectrum that arises from the axion isocurvature perturbations is $A_{\mathrm{osc}} = 0.1$.  The characteristic mass corresponding to the comoving Hubble length scale when the axion potential starts to oscillate is  \cite{dai2020gravitational}
\begin{equation}
	M_0=2.3 \times 10^{-10}\left(\frac{50\, \mu \mathrm{eV}}{m_{\rm a}}\right)^{0.51} M_{\odot} \ ,
\end{equation}
where $m_{\rm a}$ is the axion mass in $\mu \mathrm{eV}$. The function $f(\nu)$ is defined by: 
\begin{equation}\label{eq: nu times f(nu) definition}
	\nu f(\nu)=\mathfrak{A}\left(1+(q \nu)^{-\mathfrak{p} }\right)\left(\frac{\mathfrak{q} \nu}{2 \pi}\right)^{1 / 2} \exp (-\mathfrak{q} \nu / 2) \ .
\end{equation} 
\citelink{X2021}{X2021} performed numerical simulations and fitted Eq.~(\ref{eq: nu times f(nu) definition}) to their resulting mass function. They found the best-fit parameters to be $\mathfrak{A} = 0.374$, $\mathfrak{p} = 0.19$, and $\mathfrak{q} = 1.2$. Please note that these values do not correspond to the standard Sheth-Tormen mass function. Hence, we call this the modified Sheth-Tormen formalism.

To find the expression for the mass function of axion minihalos, we can rearrange Eq.~(\ref{eq: mass_function_governing_equation}) as follows: 
\begin{equation}\label{eq: mass function final expression}
	\frac{\mathrm{d}n_0}{\mathrm{d}M}(M, z) = \frac{\nu f(\nu)}{\nu} \frac{\bar{\rho}_{\rm c}}{M} \frac{\mathrm{d}\nu}{\mathrm{d}M} \ .
\end{equation}

We now need to evaluate $\mathrm{d}\nu / \mathrm{d}M$. We do this by differentiating Eq.~(\ref{eq: nu definition}) with respect to $M$ at a fixed redshift: 
\begin{equation}\label{eq: dNu / dM - first expression}
	\frac{\mathrm{d}\nu}{\mathrm{d}M} = \delta_{\rm c}^2 \left(-\frac{2}{\sigma^3}\right) \frac{\mathrm{d}\sigma}{\mathrm{d}M} \ .
\end{equation}

Next, we need to evaluate $\mathrm{d}\sigma / \mathrm{d}M$. We do this by differentiating Eq.~(\ref{eq: sigma definition}) with respect to $M$ at a fixed redshift: 
\begin{equation}\label{eq: dSigma / dM}
	\frac{\mathrm{d}\sigma}{\mathrm{d}M} = D(z) \sqrt{\frac{3 A_{\mathrm{osc}}M_0}{2 \pi^2}} \left(-\frac{1}{2}\frac{1}{M^{3/2}}\right) \ .
\end{equation}

Substituting Eqs.~(\ref{eq: sigma definition}) and (\ref{eq: dSigma / dM}) in Eq.~(\ref{eq: dNu / dM - first expression}), we get: 
\begin{equation}
	\frac{\mathrm{d}\nu}{\mathrm{d}M} = \frac{\delta_{\rm c}^2}{D^2(z)}\left(\frac{3 A_{\mathrm{osc}}M_0}{2 \pi^2}\right)^{-1} \ .
\end{equation}

Thus, we can find the value of the mass function $\mathrm{d}n_0 / \mathrm{d}M$ of axion minihalos at a given minihalo mass and redshift using Eq.~(\ref{eq: mass function final expression}).

\section{Code to evolve an orbit}
\label{app: Code to evolve an orbit}
The position and velocity of each particle at arbitrary times are evaluated using the \texttt{LBparticles} code\footnote{\url{https: //github.com/lbparticles/lbparticles}}. The code is a \texttt{Python} implementation of the high-order epicyclic approximation developed by Ref.~\cite{lyndenbell2015} with several practical improvements. Given an initial 3D position and 3D velocity and a static potential, the code computes two series of coefficients for series in $\cos(n\eta)$ and $\cos(n\chi)$, where $\eta$ and $\chi$ are fictional angles related to the particle's  angular coordinate in the potential and the time elapsed respectively, and $n$ denotes the element of the series. The orbits are not fully analytic because their properties depend on the peri- and apocenter, which must be found numerically, and the relationship between $\chi$ and $t$ must be computed numerically for each orbit. This latter relationship can be quickly constructed for arbitrary orbits by precomputing a series of integrals on a grid of $\chi$, $e$, and $k$. Here $e$ is the eccentricity of the orbit and $k$ is a closely-related quantity chosen in Ref.~\cite{lyndenbell2015} to make the 0th order version of this approximation as accurate as possible. Given an arbitrary orbit's value of $k$ and $e$, we find the closest values of $k$ and $e$ from the precomputed grid, and perform a 2D Taylor series to evaluate $t(\chi)$.

The improvements relative to Ref.~\cite{lyndenbell2015} include the following. First, several algebraic errors in the expressions for the $\cos$ series are corrected. Second, the numerical prescription for evaluating $t(\chi)$ replaced a prescription which was highly-accurate at evaluating the period of the radial oscillations of the particle, but not incredibly accurate within a single oscillation, both of which are necessary for evaluating the particle's position and velocity at arbitrary times. Third, prescriptions for the vertical oscillation of particles embedded in thin disks were added, though we do not use them in the present work.

While it is straightforward to integrate a particle's motion in a smooth central potential, \texttt{LBparticles} allows us to simply evaluate the position and velocity of the particle at any time with a low cost and a high accuracy. In contrast, numerical integration increases in expense as $t$ advances away from the time of initialization, and would require interpolation of the solution or pre-ordained evaluation points to evaluate the position at arbitrary times.

\section{The local minimum of the $\rho_* v$ against time curve does not coincide with $\dot{Z}=0$}\label{app: local minimum of rho_*v does not coincide with z_dot=0}

The term $\rho_* v$ can be plotted against time $t$. The local minima of the $\rho_* v$ curve occur when $\mathrm{d} (\rho_* v)/ \mathrm{d}t  = 0$. Thus,
\begin{equation}\label{eq: d/dt of rho_* v - part 1}
	\frac{\mathrm{d}}{\mathrm{d} t}\left(\rho_*(R, Z) v\right)=\left[\frac{\partial \rho_*}{\partial R} \dot{R}+\frac{\partial \rho_*}{\partial Z} \dot{Z}\right] v+\rho_* \dot{v}\ ,
\end{equation}
where the superscript ``." indicates derivative with respect to time. Differentiating Eq.~(\ref{eq: rho_* expression}) with respect to $R$,
\begin{align}\label{eq: partial rho_* / partial R}
	\frac{\partial \rho_*}{\partial R} & =\sum_{d=t,T} \frac{\Sigma_{\mathrm{d}, 0}}{2 Z_{\mathrm{d}}}\left(\frac{-1}{R_\mathrm{d}}\right) \exp \left(-\frac{|Z|}{Z_{\mathrm{d}}}\right) \exp \left(-\frac{R}{R_{\mathrm{d}}}\right) \nonumber\\
	& =-\frac{\rho_*}{R_\mathrm{d}}\ .
\end{align}

The second equality in Eq.~(\ref{eq: partial rho_* / partial R}) comes from making use of the definition of $\rho_*$ in Eq.~(\ref{eq: rho_* expression}). Similarly,
\begin{align}\label{eq: partial rho_* / partial z}
	\frac{\partial \rho_*}{\partial Z} & =\sum_{d=t, T} \frac{\Sigma_{\mathrm{d}, 0}}{2 Z_{\mathrm{d}}}\left( \pm \frac{1}{Z_{\mathrm{d}}}\right) \exp \left(-\frac{|Z|}{Z_{\mathrm{d}}}\right) \exp \left(-\frac{R}{R_{\mathrm{d}}}\right) \nonumber\\
	& = \pm \frac{\rho_*}{Z_{\mathrm{d}}} \ .
\end{align}

Note that we use ``$\pm$" in Eq.~(\ref{eq: partial rho_* / partial z}). The ``$+$" sign is applicable when $Z<0$ since $|Z|=-Z$ in this regime. On the other hand, the ``$-$" sign is applicable when $Z>0$ since $|Z|=Z$ in this regime.

Substituting Eqs.~(\ref{eq: partial rho_* / partial R}) and (\ref{eq: partial rho_* / partial z}) in eqn (\ref{eq: d/dt of rho_* v - part 1}),
\begin{align}\label{eq: d/dt of rho_* v - part 2}
	\frac{\mathrm{d}}{\mathrm{d} t}\left(\rho_* v\right) & =\left[-\rho_* \frac{\dot{R}}{R_{\mathrm{d}}} \pm \rho_* \frac{\dot{Z}}{Z_{\mathrm{d}}}\right] v+\rho_* \dot{v} \nonumber\\
	& =\rho_* v\left[-\frac{\dot{R}}{R_{\mathrm{d}}} \pm \frac{\dot{Z}}{Z_{\mathrm{d}}}+\frac{\dot{v}}{v}\right] \ .
\end{align}

In the R.H.S. of Eq.~(\ref{eq: d/dt of rho_* v - part 2}), $\rho_*$ and $v$ are never zero in practice. Thus, in the L.H.S., the $\rho_* v$ curve has a local minimum when the term in the square parentheses is zero. For this, it is not sufficient that $\dot{Z}=0$. There are contributions from $\dot{R}$ and $\dot{v} / v$ as well. Thus, the local minimum of the $\rho_* v$ curve doesn't exactly coincide with the instant that $\dot{Z}=0$.

\section{Converting infall redshift to look back time}\label{app: converting infall redshift to look back time}

We start with the known relation for the Hubble parameter $H$: 
\begin{align}\label{eq: Hubble parameter expression in terms of Hubble constant}
	H^2(z)&=H_0^2\left[\Omega_{\rm m}(1+z)^3+\Omega_{\rm r}(1+z)^4+\Omega_{\Lambda}\right] \nonumber\\
	\implies H(z)&=H_0\left[\Omega_{\rm m}(1+z)^3+\Omega_{\rm r}(1+z)^4+\Omega_{\Lambda}\right]^{1 / 2} \ .
\end{align}

By definition,
\begin{equation}\label{eq: hubble paramter definition}
	H=\frac{\dot{a}}{a}=\frac{1}{a} \frac{\mathrm{d} a}{\mathrm{d} t} \ .
\end{equation}

But
\begin{equation}\label{eq: scale factor formula}
	a=\frac{1}{1+z}
\end{equation}
and
\begin{equation}\label{eq: time derivative of scale factor formula}
	\frac{\mathrm{d} a}{\mathrm{d} t}=\frac{\mathrm{d} a}{\mathrm{d} z} \frac{\mathrm{d} z}{\mathrm{d} t}=-\frac{1}{(1+z)^2} \frac{\mathrm{d} z}{\mathrm{d} t} \ .
\end{equation}

Substituting Eqs.~(\ref{eq: scale factor formula}) and (\ref{eq: time derivative of scale factor formula}) in Eq.~(\ref{eq: hubble paramter definition}),
\begin{equation}
	H(z) = -\frac{1}{1+z} \frac{\mathrm{d} z}{\mathrm{d} t}
\end{equation}
\begin{equation}\label{eq: pre-integrated lookback time}
	\implies \mathrm{d}t =-\frac{1}{H(z)} \frac{1}{1+z} \mathrm{d}z \ .
\end{equation}

Integrating Eq.~(\ref{eq: pre-integrated lookback time}) from infall redshift till today
\begin{equation}\label{eq: integrated lookback time}
	\int_t^{t_0} \mathrm{d} t^\prime= - \int_{z}^0 \frac{1}{H(z^\prime)} \frac{1}{1 + z^\prime} \mathrm{d} z^\prime \ ,
\end{equation}
where $t_0$ is the time today (as measured since the big bang) and $t$ is the time corresponding to the infall redshift. But the L.H.S. of Eq.~(\ref{eq: integrated lookback time}) is just the lookback time $T$ corresponding to the infall redshift. Substituting Eq.~(\ref{eq: Hubble parameter expression in terms of Hubble constant}) into the R.H.S. of Eq.~(\ref{eq: integrated lookback time}) gives the required expression:
\begin{multline}
	T(z) = \frac{1}{H_0} \int_0^{z} \frac{1}{1+z^\prime} \\
	\times \left[\Omega_{\rm m}(1+z^\prime)^3 + \Omega_{\rm r}(1+z^\prime)^4 + \Omega_\Lambda\right]^{-1/2} \mathrm{d}z^\prime\, .
\end{multline}

\section{Distribution of energy injection parameters and concentrations}
\label{app: Distribution of energy injection parameters and concentrations}

We want to generate the joint probability density function (PDF) of $E_{\rm frac}$ and $c$ for the entire physical population of minihalos in our simulation. We first consider our two-dimensional grid points of ordered pairs of $(M_i, z_j)$ and their corresponding weight $w_{ij}$. For each grid point, we compute the concentration $c_{ij}$ that will be assigned to that grid point. As discussed in Sections~\ref{sec: Monte Carlo sampling of orbits in singular isothermal sphere} to \ref{sec: Monte Carlo Simulations to determine the stellar-disrupted mass function of minihalos}, each grid point has a random orbit which we can use to calculate the total energy injection parameter $E_{{\rm frac},ij}$ that the assigned orbit experiences during its lifetime in the galaxy, which corresponds to infall redshift $z$. Thus, to generate the joint PDF of $E_{\rm frac}$ and $c$ for the physical population of minihalos, we first create a logarithmically spaced one-dimensional array each for the possible $E_{{\rm frac},k}$ and $c_l$. We then evaluate the binned PDF as follows:
\begin{equation}
\begin{aligned}
& P_{k l} \propto \sum_{\substack{ E_{{\rm frac},k} \leq E_{{\rm frac},ij} \leq E_{{\rm frac},k}+\Delta E_{{\rm frac},k}\\ c_l \leq c_{ij} \leq c_l+\Delta c_l}} w_{i j} \\
& 
\end{aligned}
\end{equation}
where the sum is over all $i$ and $j$ values that fulfill the specified conditions. The proportionality constant can be determined so that $\sum_{k,l}P_{kl}=1.$ It should be noted that $P_{kl}$ will be grainy, the extent to which depends on the resolution of the two-dimensional $(E_{\rm frac}, c)$ bins.

\begin{figure}
        \vspace{0.1 cm}
	\includegraphics[width=\columnwidth]{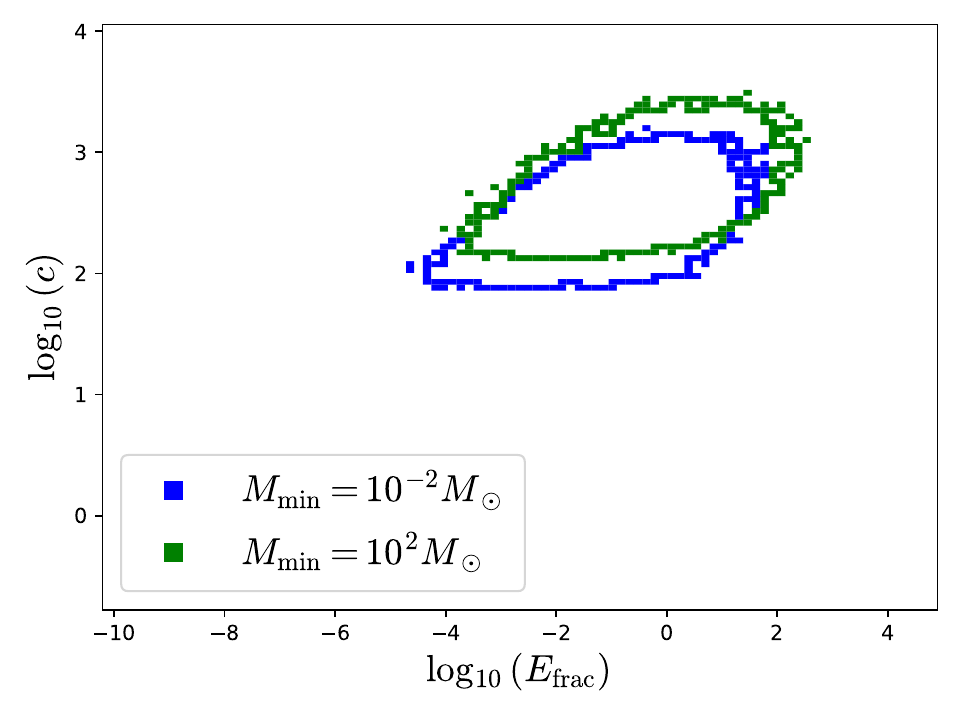}
	\caption{The  95\% contour lines of the joint probability density function (PDF) of total energy injection parameter $E_{\rm frac}$ and concentration $c$ for the entire physical population of minihalos is presented for both the $M_{\rm min}=10^{-2}M_\odot$ and $M_{\rm min}=10^{2}M_\odot$ cases. }
	\label{fig: contour plot of joint PDf of Efrac and c}
\end{figure}

 In Fig.~\ref{fig: contour plot of joint PDf of Efrac and c}, we present the  95\% confidence interval of the joint PDF of $E_{\rm frac}$ and $c$ for both the $M_{\rm min}=10^{-2}M_{\odot}$ and $M_{\rm min}=10^{2}M_{\odot}$ cases.
Each confidence interval was generated by finding the threshold value $P_{\rm thresh}$
such that 
\begin{equation}
    \sum_{P_{kl}>P_{\rm thresh}}P_{kl}\approx 0.95.
\end{equation}
for the corresponding $P_{kl}$.

In our Monte-Carlo simulation, we find that $E_{\rm frac} \in [10^{-10}, 10^4]$ and $c \in [10^{-1}, 10^4]$ approximately. Our analytical method of calculating the survival fraction given the values of $E_{\rm frac}$ and $c$ is accurate over this regime. However, as can be seen in Fig.~\ref{fig: contour plot of joint PDf of Efrac and c}, the vast majority of the physical minihalo population has $E_{\rm frac} \in [10^{-4}, 10^2]$ and $c \in [100, 3000]$ approximately. \citelink{S2024}{S2024} performed numerical simulations and generated their response function to work more or less in this narrower range. Thus, when we use \citelink{S2024}{S2024}'s response function to generate the survival fraction given $E_{\rm frac}$ and $c$ while still using the other aspects of our Monte-Carlo simulation, we find that the value of $M_{\rm surv} / M_{\rm ori} (M>10^{-12}M_\odot)$ differs from our result by (on average) 0.1\% to 1\%.

\section{Estimating the effect of updating the dynamical time after each disk pass}\label{app:updating dynamical time}

\begin{color}{black}

For simplicity, we do not update the dynamical time of the minihalo after each disk pass. To update $t_{\rm dyn}$, we would need to compute the virial radius and virial mass after each disk pass, which would be computationally intensive. In our current approach, $t_{\rm dyn}$ is set to the dynamical time at infall before any stellar disruption begins, using Eqs.~(\ref{eq: dynamical time}) and (\ref{eq:virial density}). But we now present a back-of-the-envelope calculation to estimate the effect of updating $t_{\rm dyn}$.
	
	Consider an isolated NFW profile minihalo (as considered in our previous article \cite{dsouza2024}) undergoing a stellar encounter, and then gravitationally relaxing to a Hernquist profile minihalo. For the purposes of determining the mass loss incurred by the minihalo, let our physical region of interest be the volume $V_{\mathrm{vir}, i}$ inside the virial radius of the NFW minihalo. Then, the average density $\bar{\rho}_i$ of the NFW minihalo inside volume $V_{\mathrm{vir}, i}$ is given by
	\begin{equation}\label{E1}
		\bar{\rho}_i = \frac{M_i}{V_{\mathrm{vir}, i}} ,
	\end{equation}
	where $M_i$ is effectively the virial mass of the unperturbed NFW minihalo. Next, after relaxation to the Hernquist profile, let $M_f$ be the mass of the Hernquist minihalo enclosed by the same physical volume $V_{\mathrm{vir}, i}$. 
    %Note that $M_f$ is not the virial mass of the Hernquist minihalo. 
    Then the average density $\bar{\rho}_f$ of the Hernquist minihalo inside $V_{\mathrm{vir}, i}$ is given by
	\begin{equation}\label{E2}
		\bar{\rho}_f = \frac{M_f}{V_{\mathrm{vir}, i}} .
	\end{equation}

	Dividing Eq.~(\ref{E2}) by Eq.~(\ref{E1}), we have
	\begin{equation}\label{E3}
		\frac{\bar{\rho}_f}{\bar{\rho}_i} = \frac{M_f}{M_i} .
	\end{equation}
	
	$M_f / M_i$ represents the ratio of the mass that has survived after the stellar encounter to the mass that was initially present, both evaluated inside $V_{\mathrm{vir}, i}$.
	
	Now, in Table~\ref{table:mass ratios} of our article, for the case of $p=$ hybrid, $m_{\rm a} = 25 \mu$eV, $M_{\rm min} = 10^{-2}M_\odot$, we have the ratio of the mass of the Milky Way (MW) population of minihalos that survives after stellar disruption, to the mass of the minihalo population without any stellar disruption, i.e., $M_{\rm surv}/M_{\rm ori} = 37.6\%$. In the spirit of a back-of-the-envelope calculation, we relate the isolated minihalo case to the MW minihalo population case by following the approximate equality:
	\begin{equation}\label{E4}
		\frac{M_f}{M_i} \approx \frac{M_{\rm surv}}{M_{\rm ori}}.
	\end{equation}

	Comparing Eqs.~(\ref{E3}) and (\ref{E4}), we have
	\begin{equation}\label{E5}
		\frac{\bar{\rho}_f}{\bar{\rho}_i} \approx 0.376 .
	\end{equation}
	For the MW minihalo population case, we should now think of $\bar{\rho}_i$ as an approximate measure of the dynamical time $t_{\mathrm{dyn},i}$ before any stellar disruption begins. We should also think of $\bar{\rho}_f$ as an approximate measure of the dynamical time $t_{\mathrm{dyn},f}$ of the population after all stellar encounters, i.e., today (z=0). Now, looking at Eq.~(\ref{eq: dynamical time}), we can state that
	\begin{equation}
		\frac{t_{\mathrm{dyn},f}}{t_{\mathrm{dyn},i}} = \sqrt{\frac{\bar{\rho}_i}{\bar{\rho}_f}} \approx \sqrt{\frac{1}{0.376 }} = 1.63 
	\end{equation}
	\begin{equation}
		\implies t_{\mathrm{dyn},f} \approx 1.63 \times t_{\mathrm{dyn},i}
	\end{equation}

	Now, coming back to our Monte-Carlo method, for each grid point $(M, z_{\rm i})$, we have assigned an orbit. We can say that, to a good approximation, the dynamical time at the start of stellar interactions (i.e., at $z = z_{\rm i}$) is $t_{\rm dyn}(z_{\rm i})$. On average, as time passes by, the dynamical time should increase gradually up to $1.63 \times t_{\rm dyn}(z_{\rm i})$ at the end of stellar disruptions, i.e., today $(z=0)$. Instead, if we set $t_{\rm dyn} = t_{\rm dyn}(z_{\rm i})$ and not update it in the Monte-Carlo procedure, we would be underestimating the true value of our final result: $M_{\rm surv}/M_{\rm ori}$. This is because by having a smaller $t_{\rm dyn}$, the effect of gravitational relaxation is more emphasized, leading to more mass loss, resulting in a smaller value of $M_{\rm surv}/M_{\rm ori}$ (=37.6\%). On the other hand, if we set $t_{\rm dyn} = 1.63 \times t_{\rm dyn}(z_{\rm i})$ and not update the dynamical time, we would be overestimating the value of $M_{\rm surv}/M_{\rm ori}$. We implemented this case and found out that $M_{\rm surv}/M_{\rm ori} = 38.9\%$. Thus, we conclude that by updating the dynamical time with each disk pass, it would not change the value of $M_{\rm surv}/M_{\rm ori}$ from the value presented in Table~\ref{table:mass ratios} by more than $38.9 - 37.6 = 1.3\%$.
	
	We performed a similar analysis for another case in Table~\ref{table:mass ratios}: $p=$hybrid, $m_{\rm a} = 25 \mu$eV, $M_{\rm min} = 10^2 M_\odot$. Here, we found that the value of $M_{\rm surv}/M_{\rm ori}$ wouldn't change by more than 3\% from the value presented in Table~\ref{table:mass ratios}.

\end{color}

\end{document}